\documentclass{article}
\usepackage[utf8]{inputenc}
\usepackage[legalpaper,margin=3cm]{geometry}
\usepackage{tikz}
\usetikzlibrary{shapes, arrows, positioning, automata}

\usepackage{mathpazo} 
\usepackage{pdfpages}
\usepackage{mathtools}
\usepackage{multicol}
\usepackage{graphicx}
\usepackage{amsfonts}
\usepackage{amsmath}
\usepackage[hidelinks]{hyperref}
\usepackage{amsthm}
\usepackage{listings}
\usepackage{float}
\usepackage{enumitem}
\usepackage{svg}
\usepackage{color}
\usepackage{tcolorbox}
\usepackage{fancyhdr}
\usepackage{titlesec}
\usepackage{physics}
\usepackage{caption}

\def\<#1>{\mathinner{\langle#1\rangle}}

\newcommand{\lr}[1]{\left({#1}\right)}

\newcommand{\la}{\mathcal{L}}

\newcommand{\link}[2]{{\color{blue}\href{#1}{#2}}}

\newcommand{\ls}[1]{\lstinline{#1}}

\newcommand{\vp}{\vspace{10px}}

\theoremstyle{definition}

\setlist[itemize,1]{nosep}
\setlist[enumerate,1]{nosep,label=(\alph*)}
\setlist[enumerate,2]{nosep,label=(\roman*)}

\newtcolorbox{mybox}
{
  colback=yellow!10!, colframe=red!70!black,
}

\definecolor{deepblue}{rgb}{0,0,0.8}
\definecolor{deepred}{rgb}{0.6,0,0}
\definecolor{deepgreen}{rgb}{0,0.5,0}

\DeclareFixedFont{\ttb}{T1}{txtt}{bx}{n}{12} 
\DeclareFixedFont{\ttm}{T1}{txtt}{m}{n}{12}  

\newcommand\pythonstyle{\lstset{
language=Python,
basicstyle=\ttfamily\footnotesize,
morekeywords={self},              
keywordstyle=\color{deepblue},
emphstyle=\color{deepred},    
stringstyle=\color{deepgreen},
showstringspaces=false,
}}

\lstnewenvironment{python}[1][]
{
\pythonstyle
\lstset{#1}
}
{}

\newcommand{\mytitle}{}

\title{\mytitle}
\author{Michael Sekatchev}

\begin{document}

\titlespacing\section{0pt}{2pt plus 2pt minus 2pt}{0pt plus 2pt minus 2pt}
\titlespacing\subsection{0pt}{2pt plus 2pt minus 2pt}{0pt plus 2pt minus 2pt}
\titlespacing\subsubsection{0pt}{2pt plus 2pt minus 2pt}{0pt plus 2pt minus 2pt}

\setlength{\abovedisplayskip}{0pt}
\setlength{\belowdisplayskip}{0pt}
\setlength{\abovedisplayshortskip}{0pt}
\setlength{\belowdisplayshortskip}{0pt}


\newcommand{\ou}{Ornstein-Uhlenbeck }

\begin{center}
\huge{\textbf{Stochastic approaches to asset price analysis}}
\end{center}

\tableofcontents

\pagebreak

\section{Introduction}

The late 20$^\text{th}$ century saw a rise in quantitative trading strategies, which took advantage of observed patterns and trends inferred from past behavior and current arbitrage opportunities to place profitable trades. As technology improved and the quantitative finance space grew along with it, gaining an alpha became increasingly challenging---simpler quantitative strategies, such as moving average crossover algorithms, which used to be low-hanging fruit, were now getting fully exploited and priced in, thereby requiring more and more sophisticated strategies to be devised. The field of stochastic analysis developed rapidly as a result---from Bachelier's early work treating price fluctuations as random walks, to the Black-Scholes model which revolutionized the derivatives market, this field remains at the foundation of most quantitative methods in finance to this day \cite{Bachelier}. 

An important part of this field is the forecasting of financial data. The standard methodology involves modelling a particular quantity or metric (such as asset returns \cite{forecastingTrends} or stock volatility \cite{fastStrongApproximationMCMC}) by selecting a stochastic process of choice (Geometric Brownian motion, the Ornstein-Uhlenbeck process, autoregressive moving-average (ARMA) models, mean regression models, etc.) and using a forecasting technique to estimate the model's parameters and predict the quantity's future values.  

One algorithm for implementing this is the Kalman filter \cite{kalmantutorial}. Born in the field of aerospace engineering and recently applied to the field of financial forecasting, this is an algorithm used for improving the estimates of the future states of a system: given a dynamic model of the system (such as the ones previously mentioned), the Kalman filter uses prior knowledge of the state of the system to make a prediction of its future state. At each successive timestep the algorithm compares this prediction with a realised noisy observation, obtaining a new, denoised observation. 

In this project, we propose to explore the Kalman filter's performance for estimating asset prices. We begin by introducing a stochastic mean-reverting processes, the \ou (OU) model. After this we discuss the Kalman filter in detail, and its application with this model. After a demonstration of the Kalman filter on a simulated OU process and a discussion of maximum likelihood estimation (MLE) for estimating model parameters, we apply the Kalman filter with the OU process and trailing parameter estimation to real stock market data. We finish by proposing a simple day-trading algorithm using the Kalman filter with the OU process and backtest its performance using Apple's stock price. Then we move on to a more complex model, the Heston model. This model is a combination of Geometric Brownian Motion and \ou (OU) process. There are papers like \cite{mleheston} that use MLE for parameter estimation which result in very complex forms. In our project, we propose an alternative but easier way of parameter estimation, called the method of moments (MOM). After the derivation of such estimators, we apply it to real stock data to assess its performance.

\vp 

\section{\ou Process}

In this project, we propose to explore the Kalman filter's performance for estimating asset returns. The Kalamn filter will assume that the asset returns follow a \ou process, as is popular in the field \cite{pairstrading, forecastingTrends}. 

\subsection{Basic Properties}
The OU process can be described as:
\begin{align}\label{OU process form}
    \dd X_t = \alpha (\mu - X_t) \dd t + \sigma \dd W_t
\end{align}
where $X_t$ is the price of the asset, $\alpha$ is the speed of mean reversion, and $\mu$ is a constant, indicating the expected return of the asset. 
\subsubsection{Analytic Solution}
An analytic solution can be obtained by defining a new process \( Y_t = X_t - \mu \) such that \( \dd Y_t = \dd X_t \). The original SDE (1) becomes:
\[
\dd Y_t = -\alpha Y_t \dd t + \sigma \dd W_t.
\]
Then introduce the function: \( f(t, y) = ye^{\alpha t} \).
Notice that this function has derivatives:
\[
\frac{\partial f}{\partial t} = y \alpha e^{\alpha t}, \quad \frac{\partial f}{\partial y} = e^{\alpha t}, \quad \frac{\partial^2 f}{\partial y^2} = 0.
\]
By Itô's formula:
\[
\begin{aligned}
\dd (Y_t e^{\alpha t}) &= \alpha Y_t e^{\alpha t} \dd t + e^{\alpha t} \dd Y_t \\
&= \alpha Y_t e^{\alpha t} \dd t + e^{\alpha t} \left( -\alpha Y_t \dd t + \sigma \dd W_t \right) \\
&= e^{\alpha t} \sigma \dd W_t.
\end{aligned}
\]
Integrating both sides on \([0, t]\), we have:
\[
Y_t e^{\alpha t} - Y_0 = \int_{0}^{t} e^{\alpha s} \sigma \dd W_s.
\]
By \( Y_t = X_t - \mu \), it follows that:
\begin{align}\label{anal-solution}
X_t = \mu + (X_0 - \mu)e^{-\alpha t} +
\int_0^t \sigma e^{\alpha (s-t)}\dd W_s
\end{align}
\subsubsection{Mean}
The mean/expectation of $X_t$ is: 
\begin{align*}
\mathbb{E}[X_t] &= \mathbb{E} \left[ \mu + (X_0 - \mu)e^{-\alpha t} + \int_{0}^{t} \sigma e^{\alpha(s-t)} \dd W_s \right] \\
&= \mu + (X_0 - \mu)e^{-\alpha t},    
\end{align*}
where \(\mathbb{E} \left[ \int_{0}^{t} \sigma e^{\alpha(s-t)} \dd W_s \right] = 0\) by Itô Isometry.
\subsubsection{Covariance}
The covariance function is:
\begin{align*}
\text{Cov}[X_s, X_t] &= \text{Cov} \left[ \int_{0}^{s} \sigma e^{\alpha(u-s)} \dd W_u , \int_{0}^{t} \sigma e^{\alpha(v-t)} \dd W_v \right] \\
&= \sigma^2 e^{-\alpha(s+t)} \mathbb{E} \left[ \int_{0}^{s} e^{\alpha u} \dd W_u \int_{0}^{t} e^{\alpha v} \dd W_v \right] \\
&= \sigma^2 e^{-\alpha(s+t)} \mathbb{E} \left[ \left( \int_{0}^{\min(t,s)} e^{\alpha u} \dd W_u \right)^2 + \int_{\min(t,s)}^{\max(t,s)} e^{\alpha u} \dd W_u \cdot \int_{0}^{\min(t,s)} e^{\alpha v} \dd W_v \right] \\
&= \sigma^2 e^{-\alpha(s+t)} \mathbb{E} \left[ \left( \int_{0}^{\min(t,s)} e^{\alpha u} \dd W_u \right)^2 \right] \\
&= \sigma^2 e^{-\alpha(s+t)} \int_{0}^{\min(t,s)} e^{2\alpha u} \dd u \\
&= \frac{\sigma^2}{2\alpha} \left( e^{-\alpha|t-s|} - e^{-\alpha(s+t)} \right) 
\end{align*}

\vspace{10pt}

Note that the variance of $X_t$ can be computed directly from the covariance function:
$$  \text{Var}[X_t] = \text{Cov}[X_t, X_t] = \frac{\sigma^2}{2\alpha} \left( 1 - e^{-2\alpha t} \right)$$

\vspace{10pt}

It's also worth noting that $X_t|X_{t_0}$ follows an asymptotically normal distribution; To prove this statement, we first introduce a lemma:   

\textit{Lemma}. For all adaptive functions $f(t)$, we have that: $\int_{t_0}^{t} f(s) \dd W_s \xrightarrow{D} N(0, \int_{t_0}^{t} f^2(s) ds)$. 

\textit{Proof}. WLOG, take the interval as [0, t].   

Let \( t_{n,k} = \frac{k}{2^n}t \). If follows by the definition of Itô Integral that:  
\[
\int_{0}^{t} f(s) \dd W_s = m.s\lim_{n \rightarrow \infty} \sum_{k=0}^{2^n - 1} f(t_{n,k})(W_{t_{n,k+1}} - W_{t_{n,k}}),
\]
As the increment of Brownian Motion is independent and Gaussian, we have: 
\begin{align*}
\sum_{k=0}^{2^n-1} f(t_{n,k})(W_{t_{n,k+1}} - W_{t_{n,k}}) &\sim N \left(0, \sum_{k=0}^{2^n-1} \text{Var}(f(t_{n,k})(W_{t_{n,k+1}} - W_{t_{n,k}}))\right)\\
   &= N \left(0, \sum_{k=0}^{2^n-1} f(t_{n,k})^2 2^{-n}t \right), 
\end{align*}
Since $\lim_{n\rightarrow\infty} \sum_{k=0}^{2^n-1} f(t_{n,k})^2 2^{-n}t = \int_{0}^{t} f^2(s) ds$, the lemma is proved. $\square$    

By (\ref{anal-solution}), the solution of an OU process, first notice that the function: $f(s) = \sigma e^{\alpha (s-t)}$ is indeed an adaptive function. And it follows that: 
\[
\int_{t_0}^{t} \sigma^2 e^{-2\alpha(t-s)} \, \dd s = \sigma^2 \frac{1 - e^{-2\alpha(t-t_0)}}{2\alpha}
\]
Thus by the lemma, we have: 
\begin{align}\label{ito density}
\int_{t_0}^{t} \sigma^2 e^{-2\alpha(t-s)} \xrightarrow{D} N(0, \sigma^2 \frac{1 - e^{-2\alpha(t-t_0)}}{2\alpha})
\end{align}
So by (\ref{anal-solution}), we can first rewrite the solution by setting the starting time to be $t_0$:
\[
X_t = \mu + (X_{t_0} - \mu)e^{-\alpha \left(t - t_0 \right)} +
\int_{t_0}^t \sigma e^{\alpha (s-t)}\dd W_s
\]
Hence, we have showed that:
\begin{align}\label{desitity asymptotic distribution}
X_t|X_{t_0} \xrightarrow{D} N\lr{\mu + (X_0 - \mu)e^{-\alpha \left(t - t_0 \right)}, \sigma^2 \frac{1 - e^{-2\alpha(t-t_0)}}{2\alpha}}. \quad\quad\square
\end{align}

\subsection{Numerical simulation of OU process}
There are many ways to simulate the \ou SDE, and of which two are introduced here.

\subsubsection{Using the Euler-Maruyama scheme}

The first way is through Euler-Maruyama scheme. We directly discretize the OU process \ref{OU process form}: 
\[
X_{i+1} - X_i = \alpha \left( \mu - X_i \right) \left( t_{i+1} - t_i \right) + \sigma \left(W_{i+1} - W_i \right) 
\quad \implies \quad
\implies X_{i+1} = X_i + \alpha \left( \mu - X_i \right)\Delta t_i + \sigma \Delta W_i.
\]
We divide the time interval \([0, T]\) into $0 = t_0, t_1, \ldots, t_{n-1},t_n = T$, choosing equally spaced points \( t_i \) such that \( \Delta t = t_{i+1} - t_i = \frac{T}{N} \) for each \( 1 \leq i \leq N \).    
The above formula then gets simplified to: 
\[
X_{i+1} = X_i + \alpha \left( \mu - X_i \right)\Delta t + \sigma \Delta W_i 
\]where $\space \Delta W_i \sim N\left(0, \Delta t\right)$

\vspace{10pt}

\begin{figure}[H]
    \centering
    \includegraphics[width=0.7\textwidth]{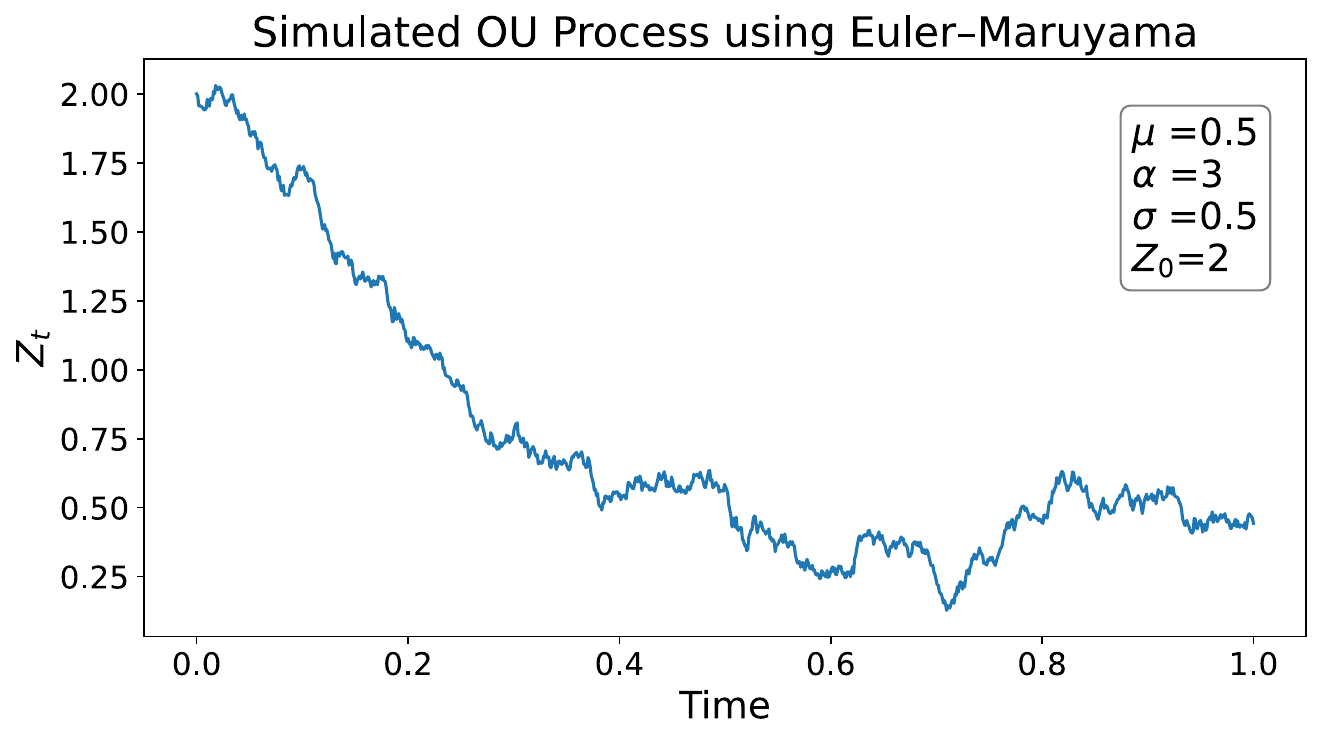}
    \caption{Simulation of \ou process using the Euler Marayama method, with the parameters shown. Notice that the process tends towards its mean of $\mu=0.5$, and favors staying in its vicinity.}
    \label{fig:ou-process}
\end{figure}

A simulated realization of an OU process using the Euler-Maruyama scheme is shown in figure \ref{fig:ou-process}. The code for this simulation can be found in \ref{sec:ou-process}.

\subsubsection{Using the analytical solution}

The second way is to directly simulate from the solution (2). We compute $X_{t+\Delta t}$ and consider the initial value at time t and by ($\ref{ito density}$) and ($\ref{desitity asymptotic distribution}$):
\begin{align}\label{discretized}
X_{t+\Delta t} = \mu + (X_t - \mu)e^{-\alpha\Delta t} + \sqrt{\frac{\sigma^2}{2\alpha} \left(1 - e^{-2\alpha\Delta t}\right)} \varepsilon_t
\end{align}
with $\epsilon_t \sim N(0, 1)$.

\vspace{10pt}

\begin{figure}[H]
    \centering
    \includegraphics[width=0.7\textwidth]{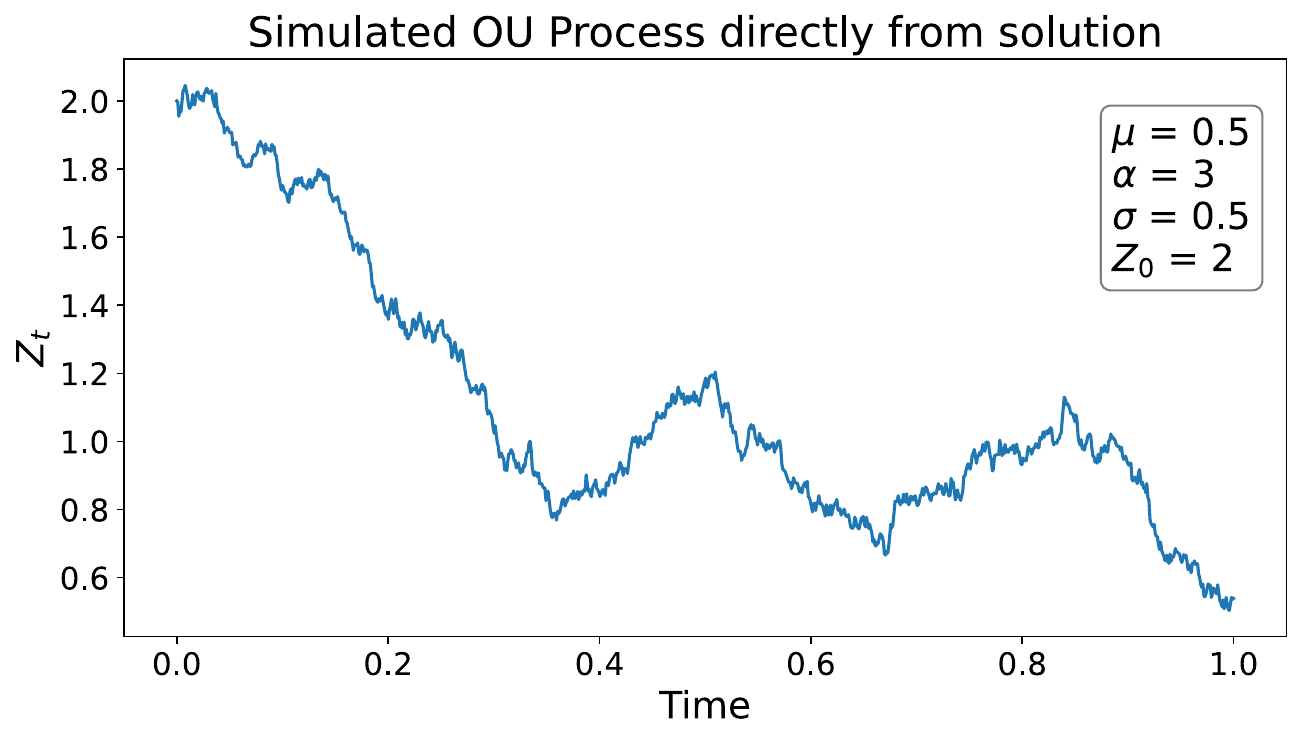}
    \caption{Simulation of \ou process directly from the solution, with the parameters shown. See Figure \ref{fig:ou-process} for comments.}
    \label{fig:ou-process-2}
\end{figure}

A simulated realization of an OU process directly from the solution is shown in figure \ref{fig:ou-process-2}. The code for this simulation can be found in \ref{sec:ou-process-2}.

\subsection{Parameter estimation using maximum log-likelihood}
\label{sec:param-estimation-ou}
Often, the optimal parameters to model a mean-reverting process are unknown, and need to be estimated. This is certainly the case with the stock price application that will be discussed in this paper. In fact, the parameters $\mu, \; \alpha, \; \sigma$ will likely evolve in time, requiring them to be re-estimated at some intervals. Two approaches for OU parameter estimation were investigated: estimating using the maximum log-likelihood, computed from the data, and estimating using the maximum log-likelihood computed after applying the Kalman filter. Out of these, the former was selected, since it was observed to produce more stable and accurate estimates. This method is discussed below.

Optimal OU parameter estimation consists of solving 
\begin{align}
\label{eqn:max-log-like}
\text{max}_{\mu,\alpha,\sigma}\la \lr{\mu,\alpha,\sigma, Z},
\end{align}
\label{eqn:log-like}
where $\la$ is the log-likelihood function given in \cite{ouparamestimation} as\begin{align}
\la\lr{\mu,\alpha,\sigma, Z}
=
-\frac{n}{2}\log\lr{\frac{\sigma^2}{2\alpha}} 
-\frac{1}{2}\sum_{i=1}^n \log\lr{1-e^{-2\alpha\Delta t_i}}
-\frac{\alpha}{\sigma^2}\sum_{i=1}^n\frac{\lr{Z_{i}-\mu-(Z_{i-1}-\mu)e^{-\alpha \Delta t_i}}^2}{1-e^{-2\alpha \Delta t_i}},
\end{align}
where $\Delta t_i = t_i - t_{i-1}$ for $i=1,...,n$. Additionally, from optimization theory, we have
\begin{align}
\label{log-like-extra-equations}
    \frac{\partial \la}{\partial \mu}=0, \quad \frac{\partial \la}{\partial (\sigma^2)} = 0, \quad \frac{\partial \la}{\partial \alpha} = 0.
\end{align}
Paper \cite{ouparamestimation} presents three different ways of solving (\ref{eqn:max-log-like}): 
\begin{enumerate}[label=\arabic*.]
    \item using a multi-dimensional numerical solver to find a set of parameters that minimizes the negative of (\ref{eqn:log-like}),
    \item solving for an optimal $\mu$ and $\sigma^2$ from the first two equations in (\ref{log-like-extra-equations}) and solving a one-dimensional minimization problem,
    \item or again using the first two equations from (\ref{log-like-extra-equations}) to find $\mu$ and $\sigma$ and then solving the last equation using a root-finder algorithm.
\end{enumerate}
The paper found that all three methods are equivalent statistically, and have a slight difference in speed (in order from fastest to slowest: 2, 3, 1). Because of this, we chose the first method for the simplicity of its implementation. In \ref{sec:ou-param-estimation}, the \ls{estimate_parameters()} function calls \ls{scipy}'s \ls{minimize} routine with the negative of (\ref{eqn:log-like}), implemented as the \ls{loglikelihood()} function.

As an example, running \ls{estimate_parameters()} on the realization shown in Figure \ref{fig:ou-process} produces parameters $(\mu, \; \alpha, \; \sigma) = (0.696, 32.7, 0.680, 0.1)$.

This function will be used in our Kalman filter implementation to recursively estimate and update the parameters of the OU model that we use for predicting asset prices. Let us now proceed to present some details on the Kalman filter, along with an application on a simulated OU process.

\vspace{10pt}

\section{Kalman Filter}
\label{sec:kalmanfiltertext}
Consider tracking the position of a car. It's true position at any point in time may follow a deterministic process, with some intrinsic noise (say, imperfect road conditions). The true position of the car may be impossible to obtain, but the position can be measured (say, using an on-board GPS). These measurements introduce some observational noise, on top of the intrinsic noise of the process itself. Given a particular model (in this case of the car's motion), The Kalman filter can use the previous mean of the process (i.e. the previous mean position of the car), and a current, noisy measurement, to estimate a new, current process mean. If the measurement is known to be very noisy and shows a strong deviation from the previous mean (say, if the GPS position sensor returns a reading off my several kilometers), the Kalman filter will more strongly rely on the model to estimate the current mean position of the car. The reverse is true if the measurement is known to be of low uncertainty. This makes the Kalman filter a very important tool for stabilizing noisy measurements of processes with high uncertainty, making it especially useful for applications in finance with highly volatile processes.

In the subsections that follow, we introduce the Kalman filter from a more quantitative perspective and apply it to an OU process with artificial Gaussian measurement noise. 

\subsection{Introduction -- a two-step process}
Let $Z_k$ be the true state of our system at some time $t_k$. The could be a 1-D scalar, or an $N$-D vector. Within the context of the Kalman filter, this true state is assumed to evolve via
\begin{align} \label{true-state}
Z_{k+1} = F_{k+1} Z_{k} + \sigma_p \epsilon_{k+1},
\end{align}
where the first term represents the deterministic evolution of the model, with $F_{k+1}$ being the state transition matrix, and where the second term represents the noise of the process, with $\sigma_p$ the standard deviation of said noise, and $\epsilon_k\sim\mathcal{N}(0,1)$. 

Suppose this true state $Z_{k+1}$ is inaccessible, but it can be measured with some Gaussian noise, obtaining an observation $\tilde{Z}_{k+1}$ through an observation model described by
\begin{align*}
\tilde{Z}_{k+1}
=
H_{k+1} Z_{k+1} + \sigma_o \epsilon'_{k+1},
\end{align*}
where $H_{k+1}$ is the observation model matrix and $\sigma_o \epsilon'_{k+1}$ is the observation noise term, with $\sigma_o$ being the standard deviation and $\epsilon'_{k+1}\sim\mathcal{N}(0,1)$.

The Kalman filter consists of estimating the mean of the true state $Z_{k+1}$, denoted $\hat{Z}_{k+1}$, given the mean of the true state in the previous timestep, $Z_{k}$, and a noisy observation of the true state at the current timestep, $\tilde{Z}_{k+1}$.
The process of obtaining this estimate can be described in two steps: a priori state prediction step, and a posteriori state update step. We will call these the prediction and update steps, as is common in literature \cite{majdaBook}. The equations that follow use the conventions from \cite{kalman-filter-main-equations}.

\subsubsection{Prediction step}
In the prediction step, an initial estimate of the mean of the true state is obtained, using only the previous mean $\hat{Z}_{k|k}$, and without using the observation at the current step. The result is denoted $\hat{Z}_{k+1|k}$, i.e. ``at the $k+1$-th step given information at $k$''. This is calculated by using the deterministic part of the true state evolution equation $(\ref{true-state})$:
\begin{align*}
\hat{Z}_{k+1|k} = F_{k+1} \hat{Z}_{k|k}.
\end{align*}
An estimate of the covariance, $P_{k+1|k}$, is also calculated via
\begin{align*}
P_{k+1|k} = F_{k+1} P_{k|k} F_{k+1}^T + Q_{k}
\end{align*}
where $Q_k$ is the covariance of the process noise. In our applications, we will be assuming that the noise at each step is independent and identically distributed (i.i.d.), so we will be setting $Q_k = \sigma_p^2\mathbb{1}$ for all $k$.

\subsubsection{Update step}
Now that we have our prior prediction of the mean, we can update it in the subsequent update step with the new noisy measurement, $\tilde{Z}_{k+1}$. To do this, the following variables are first computed:
\begin{align*}
&\text{Pre-fit residual}  & \tilde{Y}_{k+1} &= \tilde{Z}_{k+1} - H_{k+1} \hat{Z}_{k+1|k}, \\
&\text{Pre-fit residual cov.}  & S_k \;\;\; &= H_k P_{k+1|k} H_{k+1}^T + R_{k+1}, \\
&\text{Kalman gain}  & K_{k+1} &= P_{k+1|k} H_{k+1}^T S_{k+1}^{-1}.
\end{align*}
In the above, $R_k$ is the covariance of the observation noise. As with the process noise, in our applications we will assume that the noise is i.i.d., and therefore set $R_k = \sigma_o^2 \mathbb{1}$ for all $k$.

With these intermediate variables, the updated mean and covariance estimates can be computed via
\begin{align*}
&\text{Updated mean estimate}  & \hat{Z}_{k+1|k+1} &= \hat{Z}_{k+1|k} + K_{k+1}\tilde{Y}_{k+1}, \\
&\text{Updated cov. estimate}  & P_{k+1|k+1} &= (1- K_{k+1}  H_{k+1}) P_{k+1|k}.    
\end{align*}
And thus, we have obtained a new estimate of the mean of the process, which takes into account the new, noisy observation and balances it with the prediction from the model of the process.

Now, in the section that follows, we will apply these formulas to an OU process with simulated observation noise.

\subsection{Kalman filter with \ou Process}

To apply the Kalman filter on the OU process, we need to determine the state transition matrix $F_k$ as described in the section above. We begin by rearranging the discretized equation (\ref{discretized}): Writing $k$ to define the timestep at time $t_k$, and $k+1$ to denote the timestep at $t_k+\Delta t$ (where $\Delta t$ is constant for all $k$), we have
\begin{align*}
X_{k+1} = \mu + (X_k - \mu)e^{-\alpha\Delta t} + \sqrt{\frac{\sigma^2}{2\alpha} \left(1 - e^{-2\alpha\Delta t}\right)} \varepsilon_k
=
\mu\lr{1-e^{-\alpha \Delta t}} + X_k e^{-\alpha \Delta t} + \sigma_p \epsilon_k
=
A + B X_k + \sigma_p \epsilon_k,
\end{align*}
where
\begin{align*}
\sigma_p = 
\sqrt{\frac{\sigma^2}{2\alpha} \bigl( 1- e^{-2 \alpha \Delta t} \bigr)}, 
\quad
A = \mu\lr{1-e^{-\alpha \Delta t}},
\quad 
B = e^{-\alpha \Delta t}.
\end{align*}

To convert the above into the form of the Kalman Filter equations in the section above, define $Z_k = \begin{bmatrix}
    1 & X_{k}
\end{bmatrix}^T$ and rewrite the above as
\begin{align*}
Z_{k+1} = 
\begin{bmatrix}
    1 \\ X_{k+1}
\end{bmatrix}
=
\begin{bmatrix}
1 & 0 \\ A & B
\end{bmatrix}
\begin{bmatrix}
    1 \\ X_k
\end{bmatrix}
+
\begin{bmatrix}
0 \\ \sigma_p \epsilon_k
\end{bmatrix}
=
F Z_k + \begin{bmatrix}
0 \\ \sigma_p \epsilon_k
\end{bmatrix},
\end{align*}
where
\begin{align*}
F=
\begin{bmatrix}
1 & 0 \\ A & B
\end{bmatrix},
\end{align*}
a constant for all $k$. As discussed previously, assuming both the process and the observation noises are i.i.d., reduces $Q_k$ and $R_k$ to constant diagonal matrices, $Q=\sigma^2_p\mathbb{1}_2$, $R=\sigma^2_o\mathbb{1}_2$. Finally, because there is a direct mapping from the state space to the observation space, the observation matrix $H$ is simply the identity, $H=\mathbb{1}_2$.

To simulate the noisy observations, we add a new noise term to the simulated OU process shown in Figure \ref{fig:ou-process}. This term is normal with standard deviation $\sigma_o=0.1$. Following the same notation conventions, an observation at a point $k$ is given by
\begin{align*}
    \tilde{Z}_k = Z_k + \sigma_o\epsilon_k,
\end{align*}
where $\epsilon_k \sim N(0, 1)$. The true OU process with the overlayed observations is shown in Figure \ref{fig:ou-process-with-noise} below. The code for this is included in \ref{sec:ou-with-noise}.

\begin{figure}[H]
    \centering
    \includegraphics[width=0.7\textwidth]{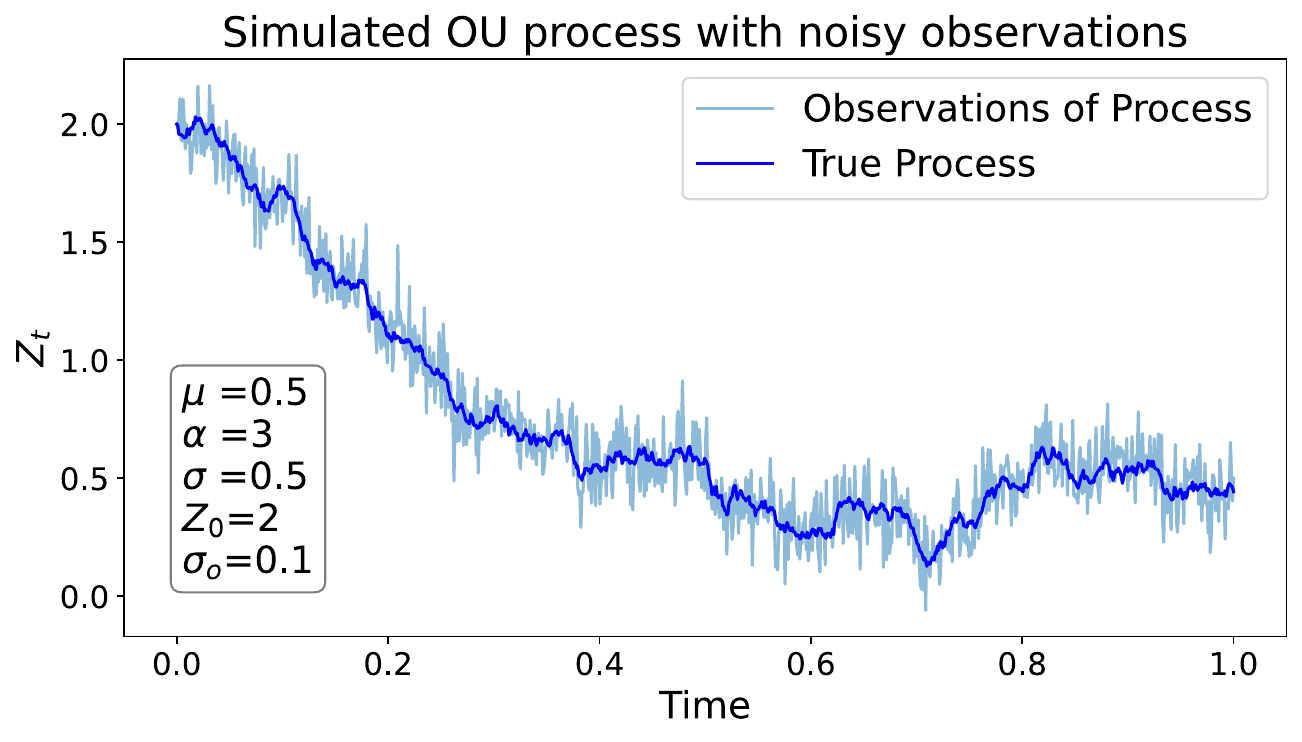}
    \caption{A simulated OU path (blue) with noisy observations of it (light blue). OU model parameters are shown, as well as the standard deviation of the noise, $\sigma_o$, which is normally distributed.}
    \label{fig:ou-process-with-noise}
\end{figure}

Having simulated the noisy measurements on top of an OU process, we can now apply the Kalman filter to these measurements to obtain a denoised estimate of the mean of the process. The Kalman filter is implemented in the \ls{kalman_filter()} function, which can be found in \ref{sec:kalman-filter-function}. For the OU process, four parameters are needed: three parameters to characterize the model, i.e. $\mu, \; \alpha, \; \sigma$, and one parameter to characterize the noise, i.e. $\sigma_o$. In this example, all of the true values were fed to the function, i.e. the same ones shown in Figure \ref{fig:ou-process-with-noise}. The result is presented in Figure \ref{fig:ou-process-with-kalman} below.

\begin{figure}[H]
    \centering
    \includegraphics[width=0.8\textwidth]{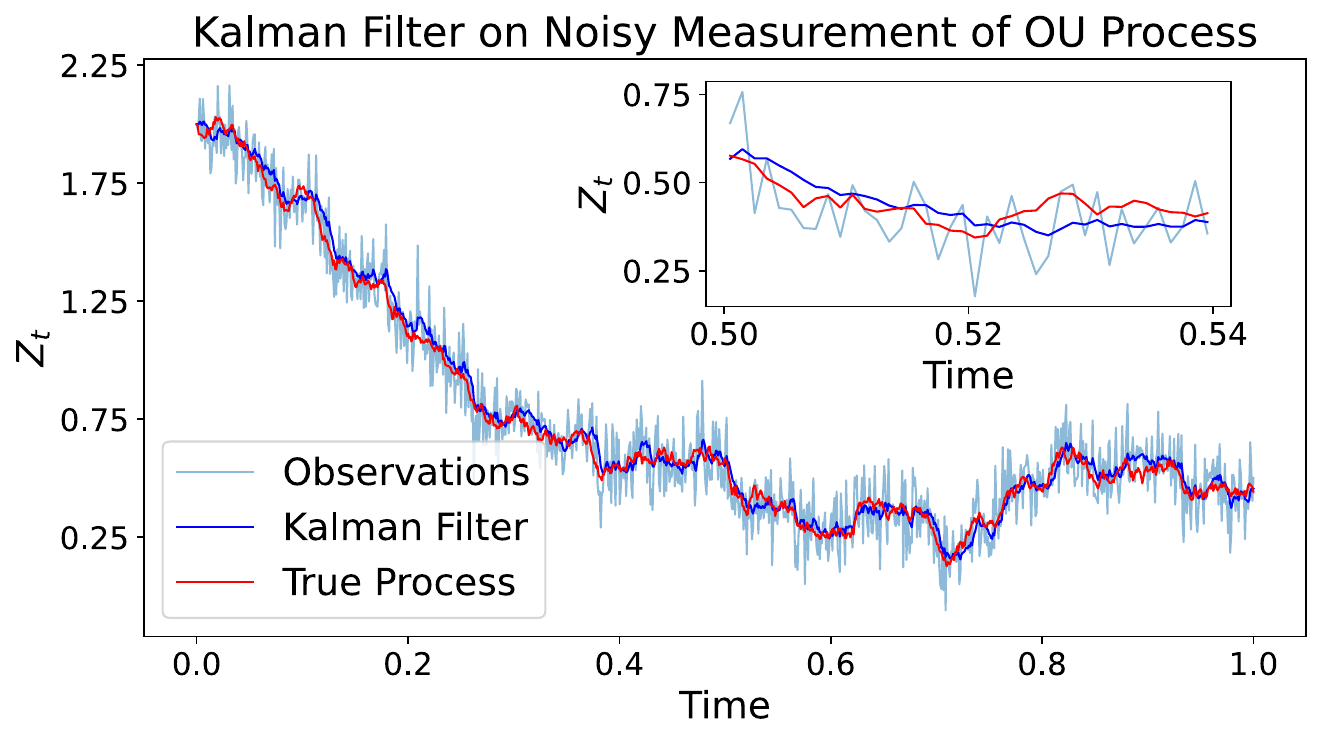}
    \caption{Running the Kalman filter on a simulated OU process with artificial observation noise. All parameters are identical to Figure \ref{fig:ou-process-with-noise}. The zoomed-in view shows that the Kalman filter successfully removes much of the observation noise, and maintains a similar trajectory to the true process.}
    \label{fig:ou-process-with-kalman}
\end{figure}



\section{Kalman filter on financial data}

Having demonstrated the Kalman filter on a typical example of a simulated OU process with noisy observations, we are now ready to discuss its application to real financial data. Although the Kalman filter has many applications in finance, including the modelling of derivative pricing such as in \cite{pairstrading}, our application will focus on asset prices, similarly to \cite{forecastingTrends}. In this section, we will first present some differences in interpretation of the Kalman filter's noisy observation process. We will then present an application of the Kalman filter on Apple's daily stock prices, and demonstrate ways to refine its predictions with recurring parameter estimation. After presenting a day-trading algorithm using the Kalman filter with an OU process, we will investigate the optimal (maximizing profits from the algorithm) observation error $\sigma_o$ and lookback time for recurring parameter estimation.

\subsection{Interpretation of observation error}

Having gone through section \ref{sec:kalmanfiltertext} and the associated OU noisy observations example, it is natural to wonder what the ``observation noise'' is in the context of asset prices. After all, we have observations of the true state with no added observational noise---all pricing data can be assumed to be accurate. Instead, we propose to shift the interpretive lens, and view the observational noise as confidence in our mean-reversion model. To provide more context, a common investment strategy is to assume that the price of an asset will tend to revert to its mean \cite{mean-reversion-paper}. This mean reversion strategy can be modelled with an OU process, with the standard parameters $\mu, \; \alpha, \; \sigma$, described in the previous sections, that are specific to an asset and to a particular time-frame. In this context of asset prices, the observational error term $\sigma_o$ can then be interpreted as the ``model confidence''. In other words, the higher the value of $\sigma_o$, the less the Kalman filter will rely on observations and the more it will rely on the mean-reversion model, resulting in adjusted mean estimates that more closely follow said model. If on the other hand $\sigma_o$ is low, the Kalman filter will favor replicating the observed asset price over the value predicted by the model. In this regard, $\sigma_o$ here can be thought of having a similar interpretation to the mean reversion coefficient $\alpha$: instead of affecting the reversion to the mean $\mu$, $\sigma_o$ affects the reversion of the Kalman filter prediction to the OU model itself. Some numerical examples of this are shown in the section that follows. From now on, $\sigma_o$ will be referred to as ``model confidence''. 

\subsection{Kalman filter with OU process on Apple's daily stock price}

Using the \ls{yfinance} library in Python\footnote{\link{https://pypi.org/project/yfinance/}{pypi.org/project/yfinance/}}, the past year of Apple's stock prices at market open and market close was extracted. In this section, we will only be using market open data. As an initial demonstration, the first month of data was used to estimate the OU parameters for the Kalamn filter, using the method described in \ref{sec:param-estimation-ou}. The code for this implementation can be found in \ref{sec:kf-apple-constant-params}. The parameter estimation returned $(\mu,\alpha,\sigma)=(171, 0.05, 0.198)$. A model confidence of $\sigma_o=20$ was chosen just for the demonstration. The result is shown in Figure \ref{fig:kf-apple-constant-params} below.

\begin{figure}[H]
    \centering
    \includegraphics[width=0.7\textwidth]{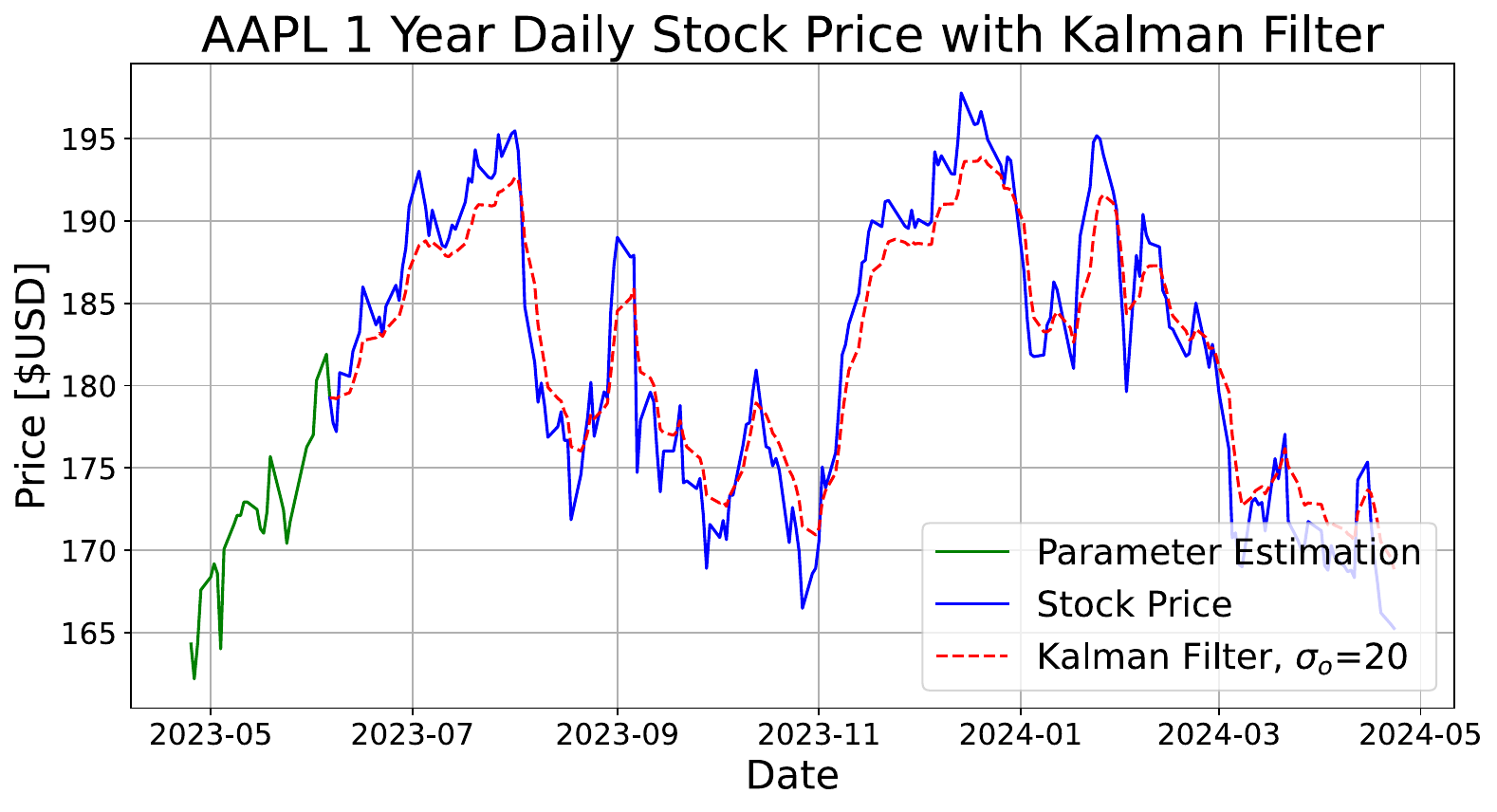}
    \caption{Kalman filter with OU process applied to Apple's daily stock market open prices for the past year. The first month of data (green) was used for OU parameter estimation for the remaining eleven months.}
    \label{fig:kf-apple-constant-params}
\end{figure}

One can notice the mean-reversion properties of the OU Kalman filter here---the filtered signal always underestimates the stock price when the latter increases sharply, and vice-versa; the filter always conservatively predicts a reversion to the mean. This is shown even more clearly in Figure \ref{fig:kf-differences} below.

\begin{figure}[H]
    \centering
    \begin{multicols}{2}
    \includegraphics[width=0.6\textwidth]{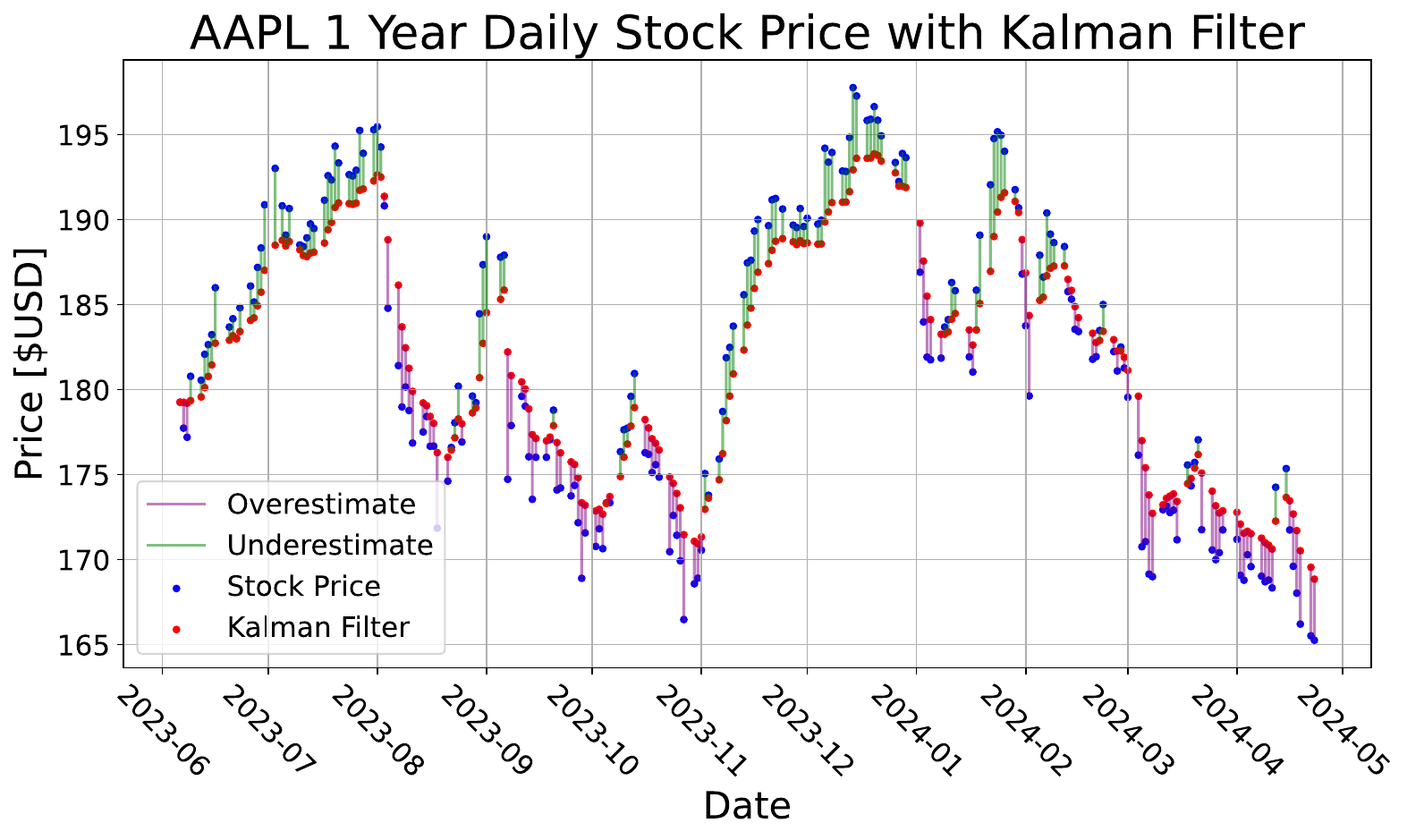}
    \columnbreak
    \includegraphics[width=0.35\textwidth]{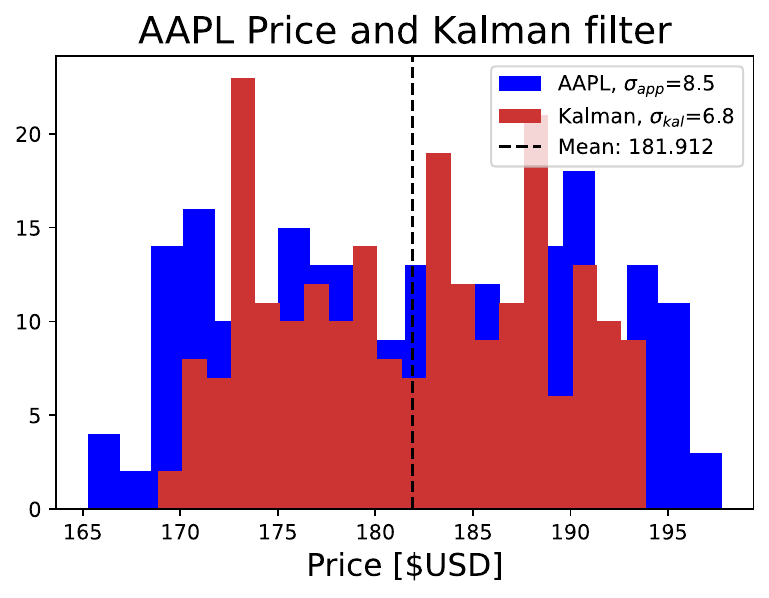}        
    \end{multicols}
    \caption{Kalman filter with OU process applied to Apple's daily stock market open prices. Left: Daily prices and Kalman filter prediction, with Kalman filter overestimates and underestimates shown in purple and green, respectively. Right: histogram of the data, with standard deviations shown. Both plots demonstrate the conservative nature of this mean-reversive OU Kalman filter process.}
    \label{fig:kf-differences}
\end{figure}

\begin{figure}[H]
    \centering
    \includegraphics[width=0.98\textwidth]{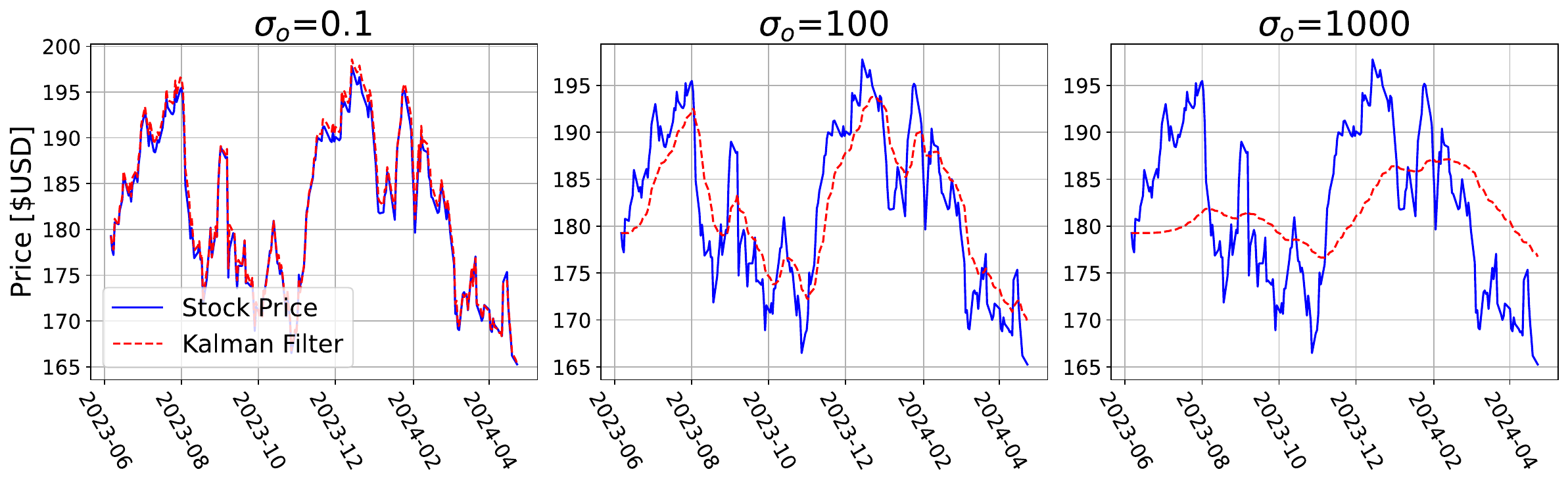}
    \caption{Apple's daily stock price in the past eleven months with the Kalman filter overlaid. Demonstrating the effect of changing the model confidence parameter, $\sigma_o$. The greater the parameter, the more the Kalman filter favors reverting to the mean of the OU process.}
    \label{fig:changing-sigma-o}
\end{figure}

The strength of this effect is influenced in part by the model confidence $\sigma_o$. Figure \ref{fig:changing-sigma-o} below demonstrates this effect clearly.

\subsection{Recursive OU parameter estimation}

As mentioned previously, the optimal OU parameters for representing a stock price are expected to evolve over time. For this reason, we propose a recursive parameter estimation method. Rather than estimating the OU parameters with a portion of the data at the start, we repeat the estimation process at each new timestep, using the previous $t_b$ days of data. This way, new information is incorporated into the model as soon as it becomes available. The implementation for this can be found in the \ls{kalman_filter_with_fitting()} function, shown in \ref{sec:kf-recursive}. This function first makes an estimate using an initial portion of the data, and then updates that estimate at every timestep once enough days for a $t_b$ backdated estimate have elapsed. A demonstration is shown in Figure \ref{fig:kf-recursive} below, with a look-back window of $t_b=30$ days, and again for a model confidence of $\sigma_o=20$.

\begin{figure}[H]
    \centering
    \includegraphics[width=0.7\textwidth]{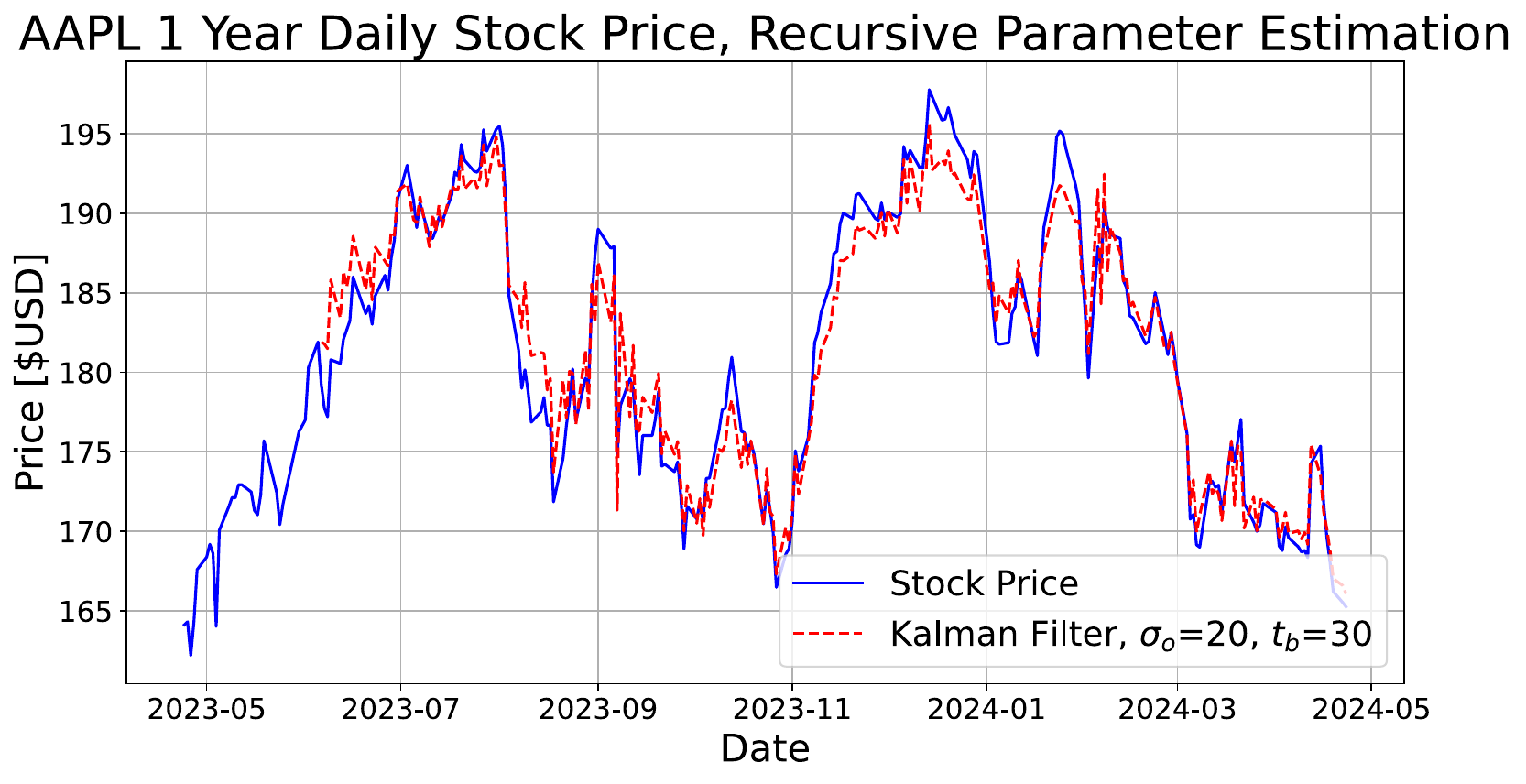}
    \caption{Apple's daily stock price in the past year with the Kalman filter overlaid. Demonstrating the effect of a recursive OU parameter estimation technique. The OU parameters are re-estimated at every timestep, with a lookback period of $t_b=30$ days.}
    \label{fig:kf-recursive}
\end{figure}
The implementation of this recursive parameter estimation method leaves us with only two parameters left to optimize/tune: the model confidence $\sigma_o$, and the lookback period $t_b$. One way to obtain these is by running backtests of historic Apple stock data, to determine which combination of $(\sigma_o, t_b)$ produces the most profit when implemented. In order to do this, we first had to design a day-trading algorithm relying on the Kalman filter estimates.

\subsection{Day-trading algorithm using Kalman filter with OU Process}

The mean-reversion predictions resulting from our OU Kalman filter have numerous applications. In this project, we decided to investigate the performance of a simple day-trading algorithm that relies on the assumption that stock prices will revert to the mean predicted by the Kalman filter. The day-trading algorithm performs the following:

At the beginning of each day, the new observation of the market open price is used with the recursive parameter estimation OU Kalman filter presented in the previous section, to produce an estimate for the open price that day. Based on the difference between the Kalman filter prediction and the true open price, a long or a short position is created with 100 shares of the stock---if the Kalman filter overestimates, then a long position is opened and vice-versa. The position is closed at the end of the day, and the profit is calculated using the closing price. All transaction costs are neglected. In the end, the final loss/gain is calculated via $P_T/(100Z^o_0)$, where $P_T$ is the (accumulated) profit at the final time $T$ and $Z^o_0$ is the initial open price of the stock on the first day of the algorithm's implementation. On each new day, $P_{i} = P_{i-1} \pm (Z^c_{i} - Z^o_{i})$, where $c$ and $o$ represent open and close prices respectively. The performance of the algorithm is compared to a buy and hold strategy, which involves simply buying 100 shares of the stock at $Z^o_0$ and selling them at time $T$. The loss/gain of this benchmark is then simply given by $(Z^c_T-Z^o_0) / Z^o_0$.

An implementation of this algorithm can be found in \ref{sec:daytradingalgo}, with a result using the same $(\sigma_o, t_b) = (20, 30)$ shown in Figure \ref{fig:kf-daytradingalgo} below.
\begin{figure}[H]
    \centering
    \includegraphics[width=0.7\textwidth]{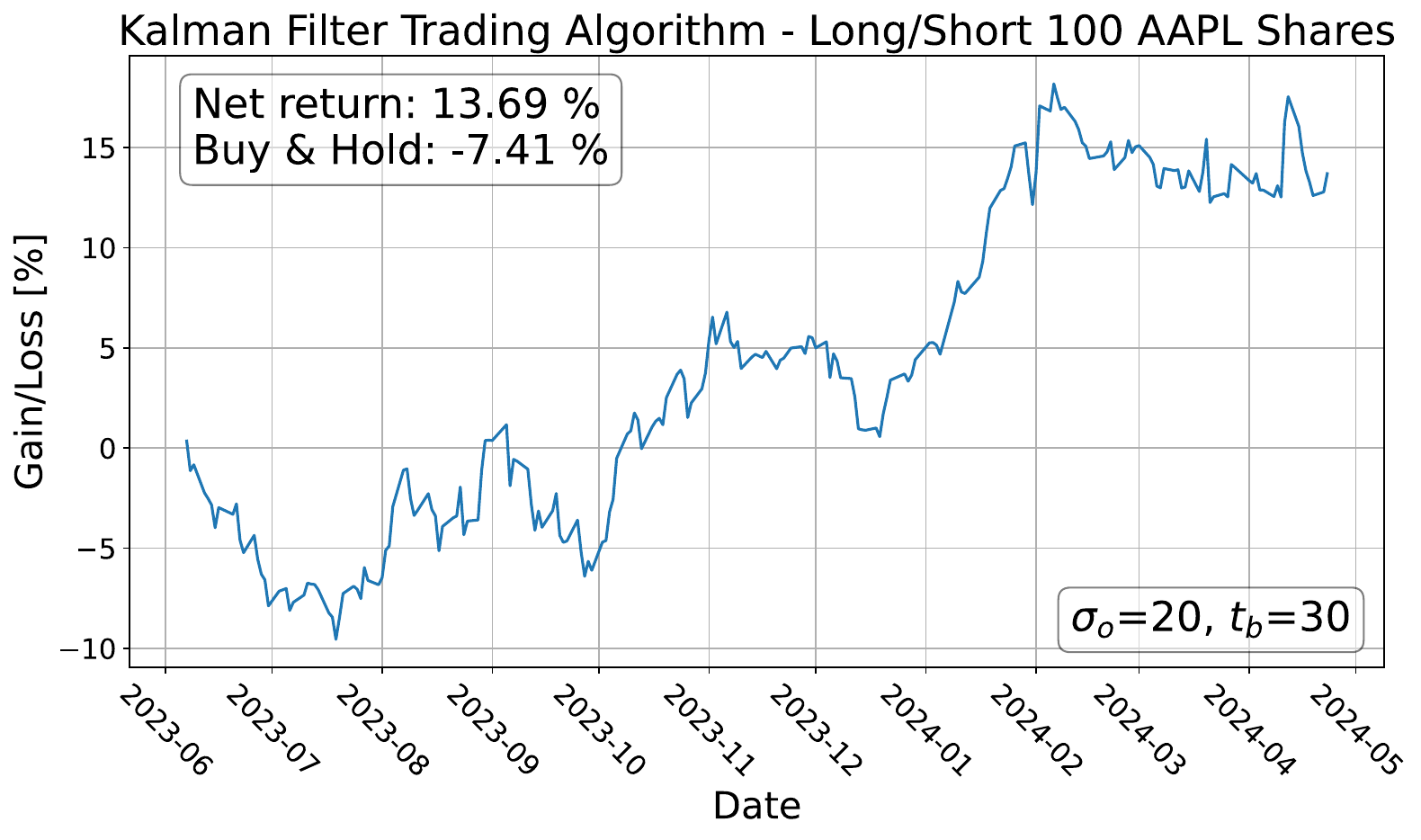}
    \caption{Implementation of the Kalman filter trading algorithm on the past year of Apple's stock price. The algorithm is seen to outperform the buy and hold strategy over this timeframe by over 21\%.}
    \label{fig:kf-daytradingalgo}
\end{figure}
Despite the success of this example with a particular set of parameters ($\sigma_o, t_b$), a robust method for determining an optimal set of parameters still needs to be created. In the next section, we perform an investigation of the ($\sigma_o, t_b$) parameter space and produce a best set of parameters estimated from backtesting on Apple's stock price in the past four years.

\subsection{Investigating optimal model confidence and lookback time}
Finally, we can perform backtesting of our trading algorithm to determine the optimal set of parameters ($\sigma_o, t_b$). To do this, a lookback period of four years was used. Of those four years, the first three were used for the backtesting (a ``training set'') and estimating an optimal set of parameters, and the last year was used for testing the algorithm based on the optimal parameters determined by the backtest. Although a more elaborate parameter estimation method could be used, we settled on running the trading algorithm for pairs of parameters in a parameter space $\sigma_o\in[10^{-2}, 10^{2}],\;t_b\in[20,200]$, similar to the process performed in \cite{forecastingTrends}. The code for the implementation is shown in \ref{sec:meshgrid-optimal-params}. A plot of the parameter space with the profits for the three-year backtesting period is shown in Figure \ref{fig:meshgrid} below.

\begin{figure}[H]
    \centering
    \includegraphics[width=0.7\textwidth]{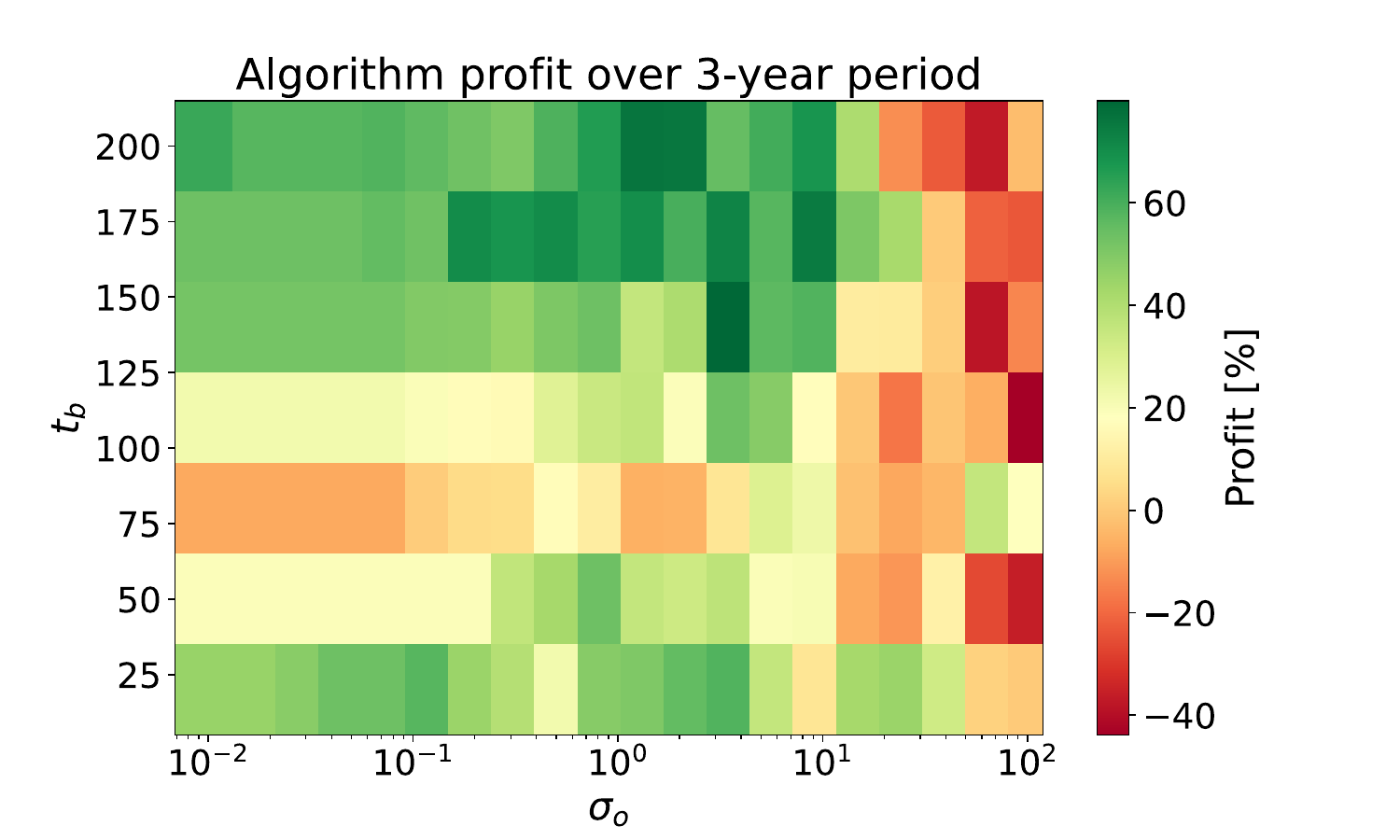}
    \caption{Percentage losses/gains obtained from running the recursively estimating OU Kalman filter on three years of Apple's stock prices using combinations of parameters ($\sigma_o, t_b$), with $\sigma_o\in[10^{-2}, 10^{2}],\;t_b\in[20,200]$. The optimal pair of parameters from this plot is $\sigma_o = 3.36,\;t_b = 140$ days.}
    \label{fig:meshgrid}
\end{figure}

We can see the limits of the parameter space as the edges are approached: to the left, for small $\sigma_o$, the model closely mimics the true price of the stock, thus limiting profits and resulting in very little change as $\sigma_o$ is decreased further. To the right, a large $\sigma_o$ results in models that deviate too strongly from the path of the stock price, quickly resulting in significant losses. From above, the backtesting period of three years limits the lookback time---a lookback time of 200 is already using almost a full year of data (a year has 252 market days). From below, a small lookback time will cause instability in the parameter estimation scheme, due to a lack of datapoints to estimate on. 

Applying the optimal set of parameters $\sigma_o = 3.36,\;t_b = 140$ to the next year of data produces a return of 12.18\%, lower than the return of 13.69\% from the example in the previous section, as shown in Figure \ref{fig:kf-daytradingalgo-optimized} below. One should also note that there are several regions in the plot which show high returns, indicating that there could certainly be more than one set of optimal parameters.

\begin{figure}[H]
    \centering
    \includegraphics[width=0.65\textwidth]{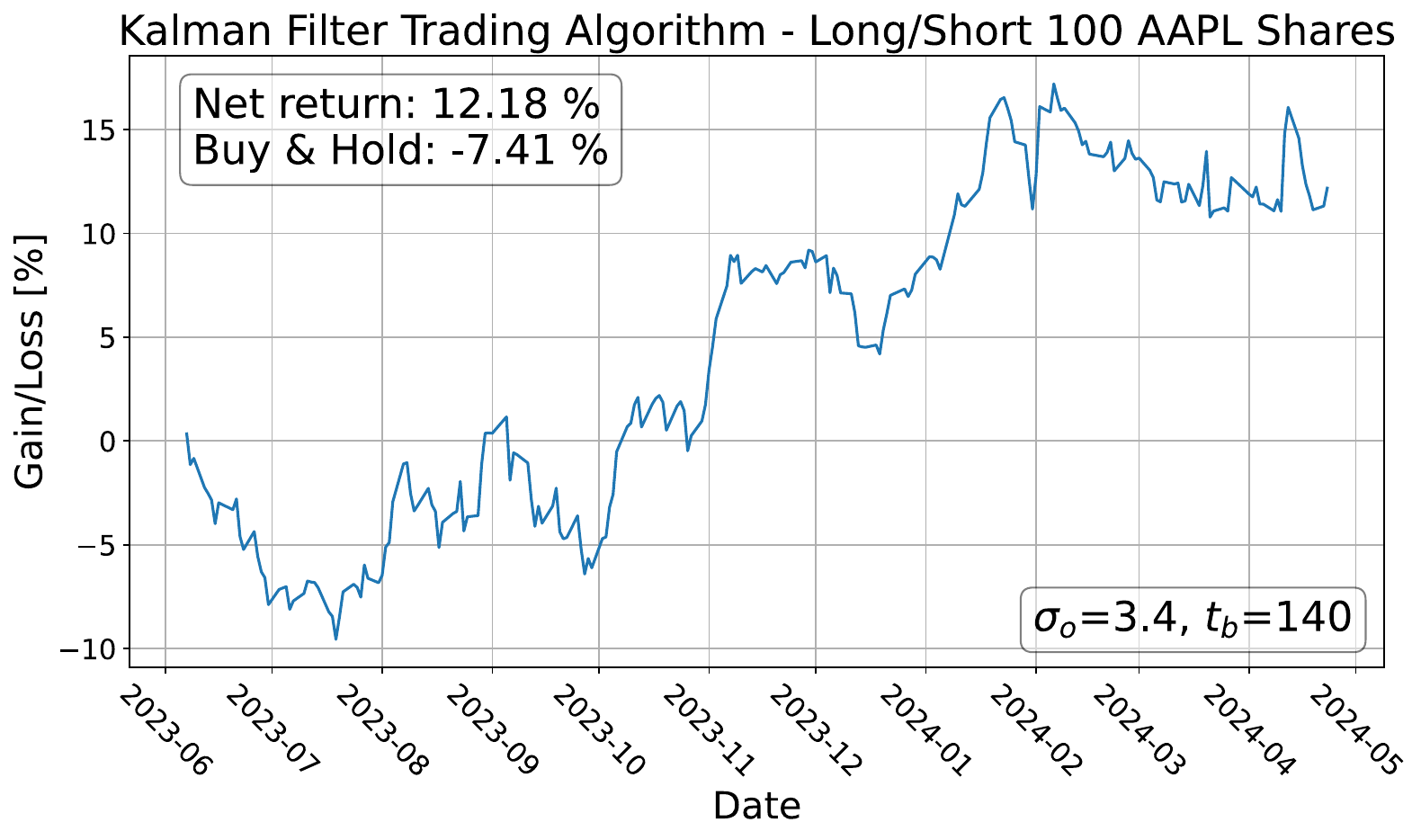}
    \caption{Implementation of the Kalman filter trading algorithm on the past year of Apple's stock price, using the optimal set of parameters determined via the backtesting method in Figure \ref{fig:meshgrid}. The profit is lower than the example shown in Figure \ref{fig:kf-daytradingalgo}.}
    \label{fig:kf-daytradingalgo-optimized}
\end{figure}

This indicates that, as expected, past performance is not evidence of future returns, and the best set of parameters for any three years of data will not necessarily be the most optimal ones to use for the subsequent year. A deeper investigation into optimal parameter estimation should be conducted---potential next steps are discussed in section \ref{sec:conclusion-and-future}. This concludes our discussion of the Kalman filter as a tool for asset price analysis.

\vspace{10pt}

\section{The Heston model}

Next, we will delve into a different, more sophisticated approach to asset pricing: the Heston model. Within the realm of mathematical finance, this model is widely used in options pricing. Despite its significance, our current level of knowledge imposes certain constraints on our exploration of its full array of characteristics. This model continues to be the subject of extensive research by numerous mathematicians and economists.

In our project, we will focus primarily on a basic simulation of the Heston model and its application to real stock data for price estimation. We propose employing the method of moments (MOM) estimator to determine the model's parameters. Our objective is to examine the implications of increasing model complexity while utilizing a comparatively ``weaker" estimation technique on the accuracy of asset price predictions.

\vspace{10pt}

\subsection{Introduction of the Heston Model}
The Heston model(shown below) is similar to the standard geometric Brownian motion apart from the fact that the volatility is a stochastic, mean-reverting process. 

\begin{align}\label{Heston model-price}
\dd S_t &= \mu S_t \dd t + \sqrt{v_t} S_t \dd W_t^S 
\end{align}

\begin{align}\label{Heston model-volatility}
\dd v_t &= \alpha (\theta - v_t) \dd t + \xi \sqrt{v_t} \dd W_t^v
\end{align}
Where:     

\(S_t\) is the stochastic process that an asset price follows

\(V_t\) is the stochastic process that volatility of the asset follows

\( \dd W_t^S \) and \( \dd W_t^v \) are Brownian with instantaneous correlation \( \rho \)

\(\mu\) = the drift term

\(\theta\) = the long-run average price variance

\(\alpha\) = the rate of mean reversion for the variance

\(\xi\) = the volatility of the volatility    

\vspace{10pt}

\subsection{Simulation of the Heston Model}
To simulate the Heston Mode, we first recognize that (\ref{Heston model-price}) is actually a geometric Brownian Motion. The solution of (\ref{Heston model-price}) is given by:
\begin{align*}
    S_t = S_0 e^{(\mu - \frac{1}{2} V_t ) t + \sqrt{V_t}W_t^S}
\end{align*} 
Assume on a uniform mesh, We compute $S_{t+\Delta t}$ and consider the initial time at t and discretize:
\begin{align} \label{simulation-Heston-Price}
    S_{t+\Delta t} = S_t e^{(\mu - \frac{1}{2} V_t) \Delta t + \sqrt{V_t}\Delta W_t^S}
\end{align}
\begin{align} \label{simulation-Heston-Volatility}
    V_{t + \Delta t} = V_t + \alpha (\theta - V_t) \Delta t + \xi \sqrt{V_t} \Delta W_t^V
\end{align}

\begin{figure}[H]
    \centering
    \includegraphics[width=0.8\textwidth]{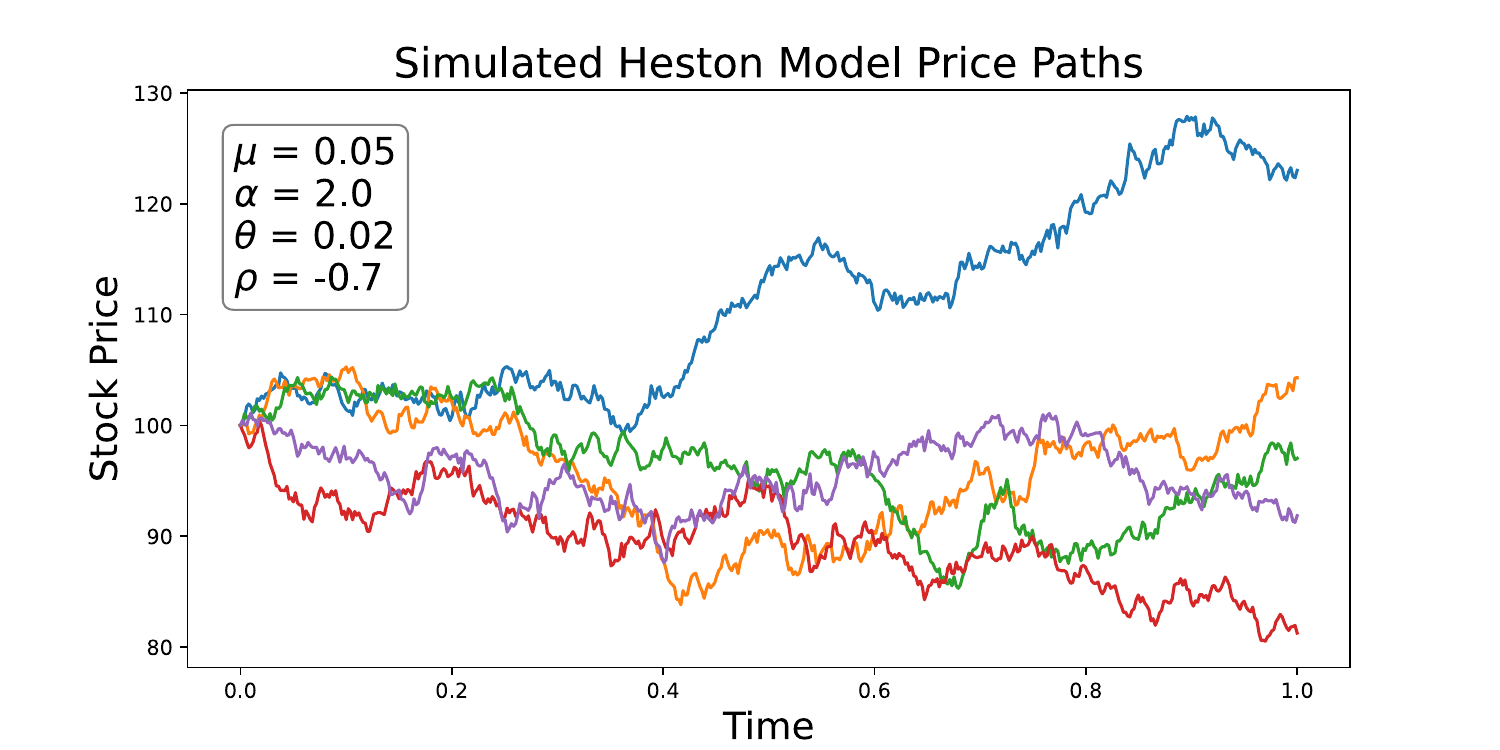}
    \caption{Five simulated Heston model stock price paths using the parameters shown.}
    \label{fig:heston-simulation}
\end{figure}

A simulated realization of 5 paths of the Heston model is shown in Figure \ref{fig:heston-simulation}. The code for this simulation can be found in \ref{sec: Heston simulation}.

\vspace{10pt}

\subsection{Method of moments for parameter estimation}
The method of moments is a common way of getting estimators of parameters. Although being not optimal, they are often easy to compute, which is especially useful in dealing with non-trivial cases like the Heston model. The formal definition can be find in the book \cite{allofstats}. 

To derive the method of moments for the Heston model, we first discretize (\ref{Heston model-price}). As in the Kalman section, we set $\dd t =$ 1 to represent one trading day between each asset price observation. The discretized form is:  

\begin{align}\label{euler-discretiztion of price}
    S_{t+1} = S_t + \mu S_t + \sqrt{V_t}S_t Z^S ,
\end{align}
where $Z^S \sim$ N(0,1).    
We denote $\frac{S_{t+1}}{S_t} = Q_t$, it follows that (\ref{euler-discretiztion of price}) is equivalent to:
\begin{align}\label{change variables: discretization of price}
    Q_{t+1} = 1 + \mu + \sqrt{V_t} Z^S
\end{align}
Next, we discretize (\ref{Heston model-volatility}). Still, we set $dt$ = 1, it follows that the discretized form is: 
\begin{align} \label{euler-discretize of volatility}
    V_{t+1} = V_t + \alpha (\theta - V_t) + \xi \sqrt{V_t}Z^V ,
\end{align}
where $Z^V \sim$ N(0,1).  

Notice that in the original model (\ref{Heston model-price}) and (\ref{Heston model-volatility}), we have $\text{Corr}$($W^S_t, W^V_t)$ = $\rho$, this property gets preserved within $Z^S$ and $Z^V$. So we set $Z^V$ = $Z_1$ and $Z^S = \rho Z_1 + \sqrt{1 - \rho^2} Z_2$, where $Z_1, Z_2 \stackrel{\text{iid}}{\sim}$ N(0,1). We do this to avoid dealing with correlated random variables, which can be a problem in deriving the method of moments estimators. This ensures that the variance of $Z^S$ is still 1 as well as transforming correlated random variables into uncorrelated ones, which is a common approach in statistics. Plug this substitution into (\ref{change variables: discretization of price}) and (\ref{euler-discretize of volatility}), we have: 
\begin{align}\label{final discretization of HM-P}
    Q_{t+1} = 1 + \mu + \sqrt{V_t} (\rho Z_1 + \sqrt{1 - \rho^2} Z_2)
\end{align}
\begin{align}\label{final discretization of HM-V}
    V_{t+1} = V_t + k(\theta - V_t) + \xi\sqrt{V_t}Z_1.
\end{align}   
Follow the recipe of \cite{allofstats}: The \(j^{\text{th}}\) moment of the random variable \(Q_{t+1}\) is defined as \(E(Q_{t+1}^j)\). We use \(m_j\) to denote the \(j^{\text{th}}\) moment.
We need to perform the following steps: 

\begin{enumerate}
    \item Write n moments in terms of the n parameters that we are trying to estimate.
    \item Obtain sample moments from the data set. The \(j^{\text{th}}\) sample moment, denoted \(\hat{m}_j\), is obtained by raising each observation to the power of \(j\) and taking the average of those terms. Symbolically,
    \[
    \hat{m}_j = \frac{1}{n} \sum_{t=1}^{n} Q_{t+1}^j.
    \]
    \item Substitute the \(j^{\text{th}}\) sample moment for the \(j^{\text{th}}\) moment in each of the n equations. That is, let \(m_j = \hat{m}_j\). Now we have a system of n equations in n unknowns.
    \item Solve for each of the n parameters. The resulting parameter values are the method of moments estimators. We denote the method of moments estimator of a parameter as \(\hat{\beta}_{MOM}\).
\end{enumerate}
We can see that there are 5 parameters: $\mu$, $\alpha$, $\theta$, $\xi$ and $\rho$ in the Heston Model (\ref{Heston model-price}) and (\ref{euler-discretize of volatility}). So theoretically, We will need to work out the first up to the fifth moment of $Q_{t+1}$ in order to get the estimators for the five parameters. The derivation of each is given below.     
\vspace{10pt}    
Derivation for the first moment $m_1$:
\[
\begin{aligned}
     m_1 &= \mathbb{E}(Q_{t+1}) = \mathbb{E}\left(1 + \mu + \sqrt{V_t} (\rho Z_1 + \sqrt{1 - \rho^2}Z_2)\right) \text{by (\ref{final discretization of HM-P})} \\
    &= 1 + \mu  \text{ since } \mathbb{E} \left( Z_1 \right) =  \mathbb{E} \left( Z_2 \right) = 0
\end{aligned}
\]    
Derivation of the second moment $m_2$:
\[
\begin{aligned}
    m_2 &= \mathbb{E}(Q_{t+1}^2) = \mathbb{E}\left(\left(1 + \mu + \sqrt{V_t} (\rho Z_1 + \sqrt{1 - \rho^2} Z_2)\right)^2\right) \\
    &= 1 + 2 \mu + \mu^2 + \mathbb{E}(V_t)\rho^2 - \mathbb{E}(V_t)\rho^2 + \mathbb{E}(V_t) \\
    &= (1+\mu)^2 + \mathbb{E}(V_t)
\end{aligned}
\]
Notice that:  
\begin{align*}
    \mathbb{E}(V_{t+1}) = \mathbb{E}(V_t) \implies \mathbb{E}(V_t) = \mathbb{E}(V_t + \alpha \left(\theta - V_t\right)) \text{ by (\ref{final discretization of HM-V}) } \implies \mathbb{E}(V_t) = \theta
\end{align*}
So it follows that: 
\begin{align*}
    m_2 = (1+\mu)^2 + \theta
\end{align*}    
Derivation of the second moment $m_3$:
\[
\begin{aligned}
    m_3 &= \mathbb{E}(Q_{t+1}^3) = \mathbb{E}\left(\left(1 + \mu + \sqrt{V_t} (\rho Z_1 + \sqrt{1 - \rho^2} Z_2)\right)^3\right) \\
    &= 1 + 3\mathbb{E}(V_t) \rho^2 - 3\mathbb{E}(V_t) \rho^2 + 3\mu + 3\mu^2 + 3\mathbb{E}(V_t) + 3\mu \rho^2 E(V_t) + \mu^3\\
    &= 1 + 3\mu + 3\mu^2 + 3\mathbb{E}(V_t) + 3\mu \mathbb{E}(V_t) + \mu^3\\
    &= (1 + \mu )^3 + 3\theta + 3\mu\theta
\end{aligned}
\]
Derivation of the fourth moment $m_4$:  
\[
\begin{aligned}
    m_4 &= \mathbb{E}(Q_{t+1}^4) = \mathbb{E}\biggl(\biggl(1 + \mu + \sqrt{V_t} (\rho Z_1 + \sqrt{1 - \rho^2} Z_2)\biggr)^4\biggr) \\ 
    &= 1 + 6\rho^2 \mathbb{E}(V_t) - 6\rho^2 \mathbb{E}(V_t) + 4\mu + 6\mu^2 \rho^2 \mathbb{E}(V_t) - 6\mu^2 \rho^2 \mathbb{E}(V_t) \\
    &\quad - 6\mathbb{E}(V_t^2) \rho^4 + 6\mathbb{E}(V_t^2) \rho^2 - 6\mathbb{E}(V_t^2) \rho^2 + \mu^4 + 6\mu^2 + 6\mathbb{E}(V_t) \\
    &\quad + 12\mu\rho^2 \mathbb{E}(V_t) - 12\mu\rho^2 \mathbb{E}(V_t) + 12\mu \mathbb{E}(V_t) + 3\mathbb{E}(V_t^2) \rho^4 \\
    &\quad + 6\mu^2 \mathbb{E}(V_t) + 3\mathbb{E}(V_t^2) + 3\mathbb{E}(V_t^2) \rho^4 + 4\mu^3
\end{aligned}
\]     
\vspace{10pt}
Here, we have to deal with the term $\mathbb{E}(V_t^2)$. Again by (\ref{final discretization of HM-V}), we have that: 
\[
\begin{aligned}
    \mathbb{E}(V_{t+1}^2) &= \mathbb{E}(V_t^2)= \mathbb{E}\left(\left(V_t + \alpha(\theta - V_t) + \xi\sqrt{V_t}Z_1\right)^2\right)\\
    &= \mathbb{E}(V_t^2) - 2\alpha\mathbb{E}(V_t^2) + 2\alpha\theta^2 + \alpha^2\theta^2 + \xi^2\theta \\
    &= \frac{-\alpha^2 \theta^2 +2\alpha \theta^2 +\xi^2 \theta}{2\alpha - \alpha^2}
\end{aligned}
\]
So it follows that:   
\begin{align*}
    m_4 = \frac{1}{\alpha(\alpha - 2)}\biggl( 
    &\alpha^2 \mu^4 + 4\alpha^2 \mu^3 - 2\alpha \mu^4 - 8\alpha \mu^3 + 6\alpha^2 \mu^2 \theta - 12\alpha \mu^2 \theta \\
    &+ 6\alpha^2 \mu^2 - 12\alpha \mu^2 + 12\alpha^2 \mu \theta - 24\alpha \mu \theta + 4\alpha^2 \mu - 8\alpha \mu \\
    &+ 3\alpha^2 \theta^2 - 6\alpha \theta^2 - 3 \xi^2 \theta + \alpha^2 - 12\alpha \theta - 2\alpha \biggr)
\end{align*}
\vspace{10pt}
Derivation of the fifth moment $m_5$:
\[
\begin{aligned}
    m_5 &= \mathbb{E}(Q_{t+1}^5) = \mathbb{E}\biggl(\biggl(1 + \mu + \sqrt{V_t} (\rho Z_1 + \sqrt{1 - \rho^2} Z_2)\biggr)^5\biggr) \\
    &= 1 + 10\rho^2 \mathbb{E}(V_t) - 10\rho^2 \mathbb{E}(V_t) + 10\rho^3 \mathbb{E}(V_t) - 10\rho^3 \rho^2 \mathbb{E}(V_t) \\
    &\quad - 30\rho^2 \mathbb{E}(V_t^2) + 15\rho^4 \mathbb{E}(V_t^2) + 15\rho^4 \mathbb{E}(V_t^2) + 5\mu + 30\rho^2 \rho^2 \mathbb{E}(V_t) \\
    &\quad - 30\rho^2 \rho^2 \mathbb{E}(V_t) - 30\rho^4 \mathbb{E}(V_t^2) + 30\rho^2 \mathbb{E}(V_t^2) - 30\rho^2 \mathbb{E}(V_t^2) + 5\mu^4 \\
    &\quad + 10\mu^2 + 10 \mathbb{E}(V_t) + \mu^5 + 10\mu^3 \mathbb{E}(V_t) + 15\mu \mathbb{E}(V_t^2) - 30\rho^4 \mathbb{E}(V_t^2) \\
    &\quad + 30\rho^2 \mathbb{E}(V_t^2) + 30\rho^2 \mathbb{E}(V_t) - 30\rho^2 \mathbb{E}(V_t) + 30\mu \mathbb{E}(V_t) + 15\rho^4 \mathbb{E}(V_t^2) \\
    &\quad + 30\mu^2 \mathbb{E}(V_t) + 15 \mathbb{E}(V_t^2) + 15\rho^4 \mathbb{E}(V_t^2) + 10\mu^3 \\
    &= \frac{1}{\alpha(\alpha - 2)} \biggl( \alpha^2 \mu^5 + 5\alpha^2 \mu^4 + 10\alpha^2 \mu^3 \theta - 2\alpha \mu^5 + 10\alpha^2 \mu^3 + 30\alpha^2 \mu^2 \theta \\
    &\quad + 15\alpha^2 \mu \theta^2 - 10\alpha \mu^4 - 20\alpha \mu^3 \theta + 10\alpha \mu^2 \theta^2 + 30\alpha \mu \theta + 15\alpha^2 \theta^2 - 20\alpha \mu^3 \\
    &\quad - 60\alpha \mu^2 \theta - 30\alpha \mu \theta^2 - 15 \xi^2 \theta + 5\alpha^2 \mu + 10\alpha^2 \theta - 20\alpha \mu^2 - 60\alpha \mu \theta \\
    &\quad - 30\alpha \theta^2 - 15 \sigma^2 \theta + \alpha^2 - 10\alpha \mu - 20\alpha \theta - 2\alpha \biggr) 
\end{aligned}
\]
Now we have the formulas up to the fifth moment, but notice that $\rho$ appears in none of them. We first put this issue aside, since we can try to find a rough approximation of the optimal $\rho$ by performing a grid search. we can also discard $m_3$, since it does not include any new parameters compared to $m_1$ and $m_2$. 
The next step is to solve the equations from real stock data: 
\begin{align} \label{mom-equation for data-final}
\left\{
\begin{aligned}
    m_1 &= \frac{1}{n}\sum_n Q_i \\
    m_2 &= \frac{1}{n}\sum_n Q_i^2\\
    m_4 &= \frac{1}{n}\sum_n Q_i^4\\
    m_5 &= \frac{1}{n}\sum_n Q_i^5
\end{aligned}
\right.
\end{align}

\vspace{10pt}

\subsection{Finding the MOM estimators for real stock data}  
In this section, we aim to find the MOM estimators using (\ref{mom-equation for data-final}) for Apple's stock using a dataset that contains observations of price of the past 3 years. As the stock data has an open price and close price, we will try to find two sets of MOM estimators. Note that the first thing we have to do is to transform the daily price observations by $Q_{t+1} = \frac{S_{t+1}}{S_t}$, then the reset should follow naturally by the result of the previous derivations.

\vspace{10pt}

For the open price, we solve for the estimators numerically via \ref{sec: code for MOM-OPEN}, it follows that: 
\begin{equation}
\begin{pmatrix}
    \hat{\mu}_{\text{MOM}} \\
    \hat{\theta}_{\text{MOM}} \\
    \hat{\alpha}_{\text{MOM}}\\
    \hat{\xi}_{\text{MOM}}
\end{pmatrix}
=
\begin{pmatrix}
    \mu_{\text{open}}\\
    \theta_{\text{open}} \\
    \alpha_{\text{open}}\\
    \xi_{\text{open}} \\
\end{pmatrix}
=
\begin{pmatrix}
    0.00047221366772620676\\
    0.0002995760762938282\\
    -0.004873456469136984\\
    -0.003476683292562815\\
\end{pmatrix}
\approx 
\begin{pmatrix}
    0.000472\\
    0.000300\\
    -0.004873\\
    -0.003477\\
\end{pmatrix}
\label{MOM-OPEN-ESTIMATOR}
\end{equation}

\vspace{10pt}

For the close price,  we solve for the estimators numerically via \ref{sec: code for MOM-CLOSE}, it follows that: 
\begin{equation}
\begin{pmatrix}
    \hat{\mu}_{\text{MOM}} \\
    \hat{\theta}_{\text{MOM}} \\
    \hat{\alpha}_{\text{MOM}}\\
    \hat{\xi}_{\text{MOM}}
\end{pmatrix}
=
\begin{pmatrix}
    \mu_{\text{close}}\\
    \theta_{\text{close}} \\
    \alpha_{\text{close}}\\
    \xi_{\text{close}} \\
\end{pmatrix}
=
\begin{pmatrix}
    0.0004457768251802108\\
    0.00028598297168747067\\
    0.1671085071227361\\
    -0.09582419900839795\\
\end{pmatrix}
\approx
\begin{pmatrix}
    0.000446\\
    0.000286\\
    0.167109\\
    -0.095824\\
\end{pmatrix}
\label{MOM-CLOSE-ESTIMATOR}
\end{equation}

\vspace{10pt}

\subsection{Estimate the stock price}
From the previous part, we have calculated the MOM estimators for the open and close price of Apple's stock price in a three-year period. We then proceed to estimate the stock price of this period using those estimators.     

Start with (\ref{final discretization of HM-V}), we take the initial condition $V_0 =$ 0 and obtain a sequence of data \{$V_t$\}. After that, set the initial condition $S_0$ to be the first price observation of the stock data. We recover $S_t$ by applying $Q_{t+1} = \frac{S_{t+1}}{S_t}$ to (\ref{final discretization of HM-P}). In the process, we perform a grid search for the approximately optimal $\rho$ on the interval [-1, 1] by comparing the sum of absolute error between the estimated $S_t$ and the true $S_t$.

For the open price estimation. We find that the optimal $\rho$ that results in the lowest sum of absolute error is $\rho =$ -1 via \ref{sec: optimal rho and sum of ae-3yr}. However, the sum of absolute error is huge, which reaches around 38453. This means that the average daily absolute error is about 50.

For the close price estimation. We find that the optimal $\rho$ that results in the lowest sum of absolute error is $\rho =$ 0.999 via \ref{sec: optimal rho and sum of ae-3yr}. However, the sum of absolute error is huge, which reaches around 90959. This means that the average daily absolute error is about 120, which is even more terrible than the result for the open price estimation.

\vspace{10pt}

\section{Conclusion and future work}
\label{sec:conclusion-and-future}

This project consisted of two components. In the first, we explored the application of the Kalman filter with the \ou process to the analysis of asset prices. We began by applying the Kalman filter on a simulated OU process with some noise. Afterwards, we discussed a new interpretation of observation noise as model confidence, in the context of asset prices. We applied the OU Kalman filter to Apple's daily stock price, and implemented a recursive optimal OU parameter estimation technique using minimum negative loglikelihood estimation. After creating a day-trading algorithm on the principle of this Kalman filter, we investigated its performance by changing the remaining Kalman filter parameters (model confidence and lookback time). We concluded that, as expected, using a set of optimal parameters from the training timeframe would not guarantee optimal performance in the testing timeframe. Further investigation is required to identify a way of reliable producing successful parameter estimates. This would include testing on different timeframes and with different assets---some less volatile assets may allow for this technique to be more effective. Overall, a proof of concept was shown: the \ou Kalman filter with recursive parameter estimation can in principle be used to estimate asset prices. Whether or not asset prices are too volatile for this needs to be investigated further. 

From the Heston model part, we can see that although the model becomes much more complex as well as the derivations are very lengthy, the MOM estimators perform terribly at estimating the stock price, causing us to cease further analysis. But this does not imply that the Heston model is a bad choice for estimating the asset price. \cite{mleheston-2} has shown that if we apply the Maximum likelihood estimator (MLE) on the Heston model, we can in fact achieve an accurate, or to say satisfying estimation of the asset price. We think that there are two main reasons for this. The first is from a statistical aspect: The MLE achieves the Cramér-Rao Lower bound(CRLB) under certain regularity conditions, indicating that it's asymptotically efficient and hence has the lowest variance among all unbiased estimators. While the MOM does not, it has a high possibility to miss critical aspects of the data structure especially when dealing with non-linear cases, thus causing high variance. The second is from the nature of the Heston model: This model involves stochastic processes with features like volatility clustering and mean reversion, which are modeled by complex dynamics involving multiple parameters linked in non-linear equations. The MLE can handle these complexities more effectively because it maximizes the likelihood function that accurately represents the stochastic nature of the data. While the MOM might not effectively capture these dynamics as it relies on simple and direct moment conditions, which may not sufficiently reflect the intricate dependencies between model parameters and data characteristics. So for situations involving large volatility like the stock market, the choice of estimators plays a more deterministic role in arriving at an accurate estimation.

Overall, we see that stochastic analysis of financial assets is complicated by the high volatility of the underlying processes, rendering parameter estimation a very interesting challenge.

\pagebreak

\bibliographystyle{plain}
\bibliography{biblio}

\pagebreak

\setcounter{section}{0}

\renewcommand{\thesection}{\Alph{section}} 

\section{Appendix}

The code for this project was written in Python. It requires the following packages:
\begin{python}
import numpy as np
import matplotlib.pyplot as plt
import scipy.stats as ss
from scipy.optimize import minimize
import scipy    
import yfinance as yf
\end{python}

The full version of the code can be found in the GitHub repository\footnote{\link{https://github.com/msekatchev/kalman-filter-asset-prices}{github.com/msekatchev/kalman-filter-asset-prices}}.

\subsection{Simulation of \ou Process using EM scheme}
\label{sec:ou-process}
\begin{python}
# simulate an OU process using the EM method
np.random.seed(seed=306)

# model parameters
alpha = 3
mu = 0.5
sigma = 0.5
Z0 = 2  

# simulation parameters
T = 1 
N = 1000
dt = T / N     
time = np.linspace(0, T, N)
Zt = np.zeros(N)

Zt[0] = Z0

# EM steps
for t in range(1, N):
    dW = np.random.normal(0,np.sqrt(dt)) 
    Zt[t] = Zt[t-1] + alpha * (mu - Zt[t-1]) * dt + sigma * dW

plt.figure(figsize=(10, 5), dpi=300)
plt.plot(time, Zt)
plt.title('Simulated OU Process using Euler-Maruyama', size=20)
plt.xlabel('Time', size=18)
plt.ylabel(r'$Z_t$', size=18)
plt.xticks(fontsize=15)
plt.yticks(fontsize=15)
plt.text(0.85, 0.9, r'$\mu$ ='+str(mu)+"\n"+r'$\alpha$ ='+str(alpha)+"\n"+r'$\sigma$ ='+str(sigma)+"\n"+r'$Z_0$='+str(Z0), transform=plt.gca().transAxes, fontsize=18,
         verticalalignment='top', bbox=dict(boxstyle='round', facecolor='white', alpha=0.5))
plt.savefig("ou-process.svg",bbox_inches='tight')
plt.show()
\end{python}

\subsection{Simulation of \ou Process from solution}
\label{sec:ou-process-2}
\begin{python}
np.random.seed(seed=605)

# Model parameters
alpha = 3
mu = 0.5
sigma = 0.5
Z0 = 2

# Simulation parameters
T = 1
N = 1000
dt = T / N
time = np.linspace(0, T, N)
Zt = np.zeros(N)
Zt[0] = Z0

# Simulate directly from solution
for t in range(1, N):
    Zt[t] = mu + (Zt[t - 1] - mu) * np.exp(-alpha * dt) + np.sqrt(sigma**2 / (2 * alpha) * (1 - np.exp(-2 * alpha * dt))) * np.random.normal(0, 1)

plt.figure(figsize=(10, 5), dpi=300)
plt.plot(time, Zt)
plt.title("Simulated OU Process directly from solution", size=20)
plt.xlabel("Time", size=18)
plt.ylabel(r'$Z_t$', size=18)
plt.xticks(fontsize=15)
plt.yticks(fontsize=15)
plt.text(0.85, 0.9, r'$\mu$ = ' + str(mu) + '\n' + r'$\alpha$ = ' + str(alpha) + '\n' + r'$\sigma$ = ' + str(sigma) + '\n' + r'$Z_0$ = ' + str(Z0),
         transform=plt.gca().transAxes, fontsize=18, verticalalignment='top', bbox=dict(boxstyle='round', facecolor='white', alpha=0.5))
plt.savefig("ou-process-2.svg", bbox_inches="tight")
plt.show()
\end{python}

\subsection{OU parameter estimation using minimum negative log-likelihood}
\label{sec:ou-param-estimation}
\begin{python}
def loglikelihood(c, Xt_partial, time_partial):
    mu, alpha, sigma = c
    n = len(time_partial)
    dt = time[1] - time[0]

    L = -n/2 * np.log(sigma**2 / (2*alpha)) \
        -1/2 * np.sum(np.log(1-np.exp(-2*alpha*dt))) \
        -alpha/sigma**2 * np.sum((Xt_partial[1:n] - mu - (Xt_partial[0:n-1] - mu)*np.exp(-alpha * dt))**2/(1-np.exp(-2*alpha*dt)))
    
    return -L

def estimate_parameters(x0, Xt_partial, time_partial):
    mu, alpha, sigma = x0
    result = minimize(
        loglikelihood,
        x0=[mu, alpha, sigma],
        args=(Xt_partial, time_partial),
        method = "L-BFGS-B", 
        bounds = ((None, None), (0.05, None), (0.05, None)))

    mu, alpha, sigma = result.x
    
    return mu, alpha, sigma
\end{python}

\subsection{Simulation of OU process with noisy observations}
\label{sec:ou-with-noise}
\begin{python}
# add some gaussian noise on top of the model to simulate observations
np.random.seed(seed=42)
sigma_o = 0.1
eps = ss.norm.rvs(loc=0, scale=sigma_o, size=N)
eps[0] = 0
Xt = Zt + eps  # process + noise = measurement process

plt.figure(figsize=(10, 5), dpi=300)
plt.plot(time, Xt, "-", alpha=0.5, label="Observations of Process")
plt.plot(time, Zt, "-b", label="True Process")
plt.title('Simulated OU process with noisy observations', size=20)
plt.xlabel('Time', size=18)
plt.ylabel(r'$Z_t$', size=18)
plt.xticks(fontsize=15)
plt.yticks(fontsize=15)
plt.legend(fontsize=18)
plt.text(0.05, 0.45, 
         r'$\mu$ ='+str(mu)+"\n"+r'$\alpha$ ='+str(alpha)+"\n"+r'$\sigma$ ='+str(sigma)+\
         "\n"+r'$Z_0$='+str(Z0)+"\n"+r'$\sigma_o$='+str(sigma_o), transform=plt.gca().transAxes, fontsize=18,
         verticalalignment='top', bbox=dict(boxstyle='round', facecolor='white', alpha=0.5))
plt.savefig("ou-process-with-noise.svg",bbox_inches='tight')
plt.show()
\end{python}

\subsection{OU Kalman Filter}
\label{sec:kalman-filter-function}
\begin{python}
def kalman_filter(Xt, time, mu, alpha, sigma, sigma_o):   
    # assume time intervals are uniformal
    dt = time[1] - time[0]
    
    # calculate OU model parameters for Kalman filter
    A = mu * (1-np.exp(-alpha * dt))
    B = np.exp(-alpha * dt)
    F = np.array([[1,0],[A,B]])
    sigma_p = np.sqrt(sigma**2 / (2*alpha) * (1-np.exp(-2*alpha*dt)))

    P = np.eye(2) * sigma_p**2
    H = np.eye(2)
    Q = np.eye(2) * sigma_p**2
    R = np.eye(2) * sigma_o**2

    # vector for storing Kalman predictions
    Z = np.zeros([len(Xt),2])
    # vector for residuals
    Y = np.zeros([1,2])
    # vector for noisy, measured data
    Xt = np.column_stack((np.ones_like(Xt), Xt))

    # initialize Kalman predictions vector and -loglikelihood variable
    Z[0] = Xt[0]
    L = 0

    # Kalman filter - loop through time
    for i in range(len(time)-1):
        # prediction step
        Z[i+1] = F @ Z[i]
        P = F @ P @ F.T + Q

        # update step
        Y = Xt[i+1] - H @ Z[i+1] # pre-fit residual mean
        S = H @ P @ H.T + R # pre-fit residual covariance
        K = P @ H.T @ np.linalg.inv(S) # Kalman gain
        Z[i+1] = Z[i] + K @ Y 
        P = (np.eye(2) - K @ H) @ P # post-fit covariance estimate

        # compute marginal negative loglikelihood
        # L += -1/2 * (Y.T @ np.linalg.inv(S) @ Y + np.log(np.linalg.det(S)) + 2 * np.log(2*np.pi))
    return Z
\end{python}

\subsection{Kalman filter with OU process on Apple's stock returns, fixed parameters}
\label{sec:kf-apple-constant-params}
\begin{python}
# load in QQQ data and retrieve open and close prices
qqq = yf.Ticker("AAPL")
data = qqq.history(period="1y")

open_prices = data['Open'].dropna()
close_prices = data['Close'].dropna()

open_returns = open_prices.pct_change().dropna().values

# estimate parameters on the first 30 days of data
start_index = 30

start_data = open_prices[0:start_index]

dt = 1

mu, alpha, sigma = estimate_parameters([170, 3, 0.1], start_data, np.arange(0,len(start_data),dt))

print(mu, alpha, sigma, sigma_o)

Yt = open_prices[start_index:]
time = np.arange(0,len(Yt),dt)

sigma_o = 20
Z = kalman_filter(Yt, time, mu, alpha, sigma, sigma_o)

plt.figure(figsize=(12, 6), dpi=300)
plt.plot(np.array(data.index[1:start_index+1]), 
         np.array(open_prices)[1:start_index+1],
         label=r'Parameter Estimation', color='green')
plt.plot(np.array(data.index[start_index:]), 
         np.array(open_prices)[start_index:],
         "-",label='Stock Price', color='blue')

plt.plot(np.array(data.index[start_index:]), 
         Z[:,1], "--",
         label='Kalman Filter, '+r'$\sigma_o$='+str(sigma_o), color="red")

plt.title('AAPL 1 Year Daily Stock Price with Kalman Filter', size=25)
plt.xlabel('Date', size=20)
plt.ylabel('Price [$USD]', size=20) # $
plt.xticks(fontsize=15)
plt.yticks(fontsize=15)
plt.legend(fontsize=18, loc="lower right")
plt.grid(True)
plt.savefig("kf-apple-constant-parameters.svg", bbox_inches='tight')
plt.show()
\end{python}

\subsection{Kalman filter with recursive parameter estimation}
\label{sec:kf-recursive}
\begin{python}
def kalman_filter_with_fitting(Xt, time, x0, sigma_o, start_index, lookback_period):    
    # estimate initial set of parameters using the starting data
    mu, alpha, sigma = estimate_parameters(x0, Xt[0:start_index], time[0:start_index])

    # assume time intervals are uniform
    dt = time[1] - time[0]

    # calculate OU model parameters for Kalman filter
    A = mu * (1-np.exp(-alpha * dt))
    B = np.exp(-alpha * dt)
    F = np.array([[1,0],[A,B]])
    sigma_p = np.sqrt(sigma**2 / (2*alpha) * (1-np.exp(-2*alpha*dt)))
    
    P = np.eye(2) * sigma_p**2
    H = np.eye(2)
    Q = np.eye(2) * sigma_p**2
    R = np.eye(2) * sigma_o**2

    # vector for storing Kalman predictions
    Z = np.zeros([len(Xt),2])
    # vector for residuals
    Y = np.zeros([1,2])
    # vector for noisy, measured data
    Xt = np.column_stack((np.ones_like(Xt), Xt))

    # initialize Kalman predictions vector
    Z[0:start_index] = Xt[0:start_index]

    # Kalman filter - loop through time
    for i in np.arange(start_index-1, len(time)-1):
        # prediction step
        # option to use true state here instead of previous prediction
        Z[i+1] = F @ Z[i]
        P = F @ P @ F.T + Q

        # update step
        Y = Xt[i+1] - H @ Z[i+1] # measurement pre-fit residual
        S = H @ P @ H.T + R
        K = P @ H.T @ np.linalg.inv(S) # Kalman gain
        Z[i+1] = Z[i] + K @ Y
        P = (np.eye(2) - K @ H) @ P
        
        # if more steps that the lookback_period have passed, start estimating parameters
        if i > lookback_period:
            Xt_partial = Xt[i-lookback_period:i,1]
            time_partial = time[i-lookback_period:i]
        
            mu, alpha, sigma = estimate_parameters([mu, alpha, sigma], Xt_partial, time_partial)
            
            P0 = np.eye(2) * sigma_p
            H = np.eye(2)
            Q = np.eye(2) * sigma_p
            R = np.eye(2) * sigma_o
            A = mu * (1-np.exp(-alpha * dt))
            B = np.exp(-alpha * dt)
            F = np.array([[1,0],[A,B]])
            sigma_p = np.sqrt(sigma**2 / (2*alpha) * (1-np.exp(-2*alpha*dt)))

    return Z
\end{python}

\subsection{Day-trading algorithm using the Kalman filter with recursive parameter estimation}
\label{sec:daytradingalgo}
\begin{python}
# implementing trading algorithm
close_prices
profit = 0
profit_list = []
start_price = open_prices[start_index] # + open_prices[start_index] *np.cumsum(Z[:, 1])
capital = start_price * 100
for i in range(len(open_prices.index[start_index+1:])):
    date = open_prices.index[start_index+i+1]

    open_price = open_prices[start_index+i+1]
    close_price = close_prices[start_index+i+1]
    
    pred_return = Z[i,1]
    open_return = open_returns[start_index+i]
    open_return = open_prices[start_index+i]
#     print(open_return, pred_return)
    if open_return < pred_return:
        profit += (close_price - open_price)*100
    else:
        profit += (open_price - close_price)*100
    profit_list.append(profit)


plt.figure(figsize=(12, 6), dpi=300)

plt.title('Kalman Filter Trading Algorithm - Long/Short 100 AAPL Shares', size=22)
plt.xlabel('Date', size=20)
plt.ylabel('Profit [$USD]', size=20)
plt.xticks(fontsize=15, rotation=-45)
plt.yticks(fontsize=15)
plt.grid(True)

returns = profit / capital * 100

plt.plot(np.array(open_prices.index)[start_index+1:], profit_list)
plt.text(0.05, 0.95, r'Net return: '+str(round(returns,2))+" %" + \
         "\nBuy & Hold: "+str(round((close_prices[len(close_prices)-1]-start_price)/(close_prices[len(close_prices)-1])*100,2))+" %", 
         transform=plt.gca().transAxes, fontsize=22,
         verticalalignment='top', bbox=dict(boxstyle='round', facecolor='white', alpha=0.5))
plt.text(0.75, 0.11, r'$\sigma_o$='+str(sigma_o)+r', $t_b$='+str(lookback_period), 
         transform=plt.gca().transAxes, fontsize=22,
         verticalalignment='top', bbox=dict(boxstyle='round', facecolor='white', alpha=0.5))

plt.savefig("kalman-algorithm.svg", bbox_inches='tight')

plt.show()
\end{python}

\subsection{Backtesting to determine optimal model confidence and lookback period}
\label{sec:meshgrid-optimal-params}
\begin{python}
qqq = yf.Ticker("AAPL")
data = qqq.history(period="4y")

open_prices = data['Open'].dropna()
close_prices = data['Close'].dropna()

start_index = 30

def profit_function(c, open_prices, close_prices, start_index):
    sigma_o, t_b = c
    Yt = open_prices
    time = np.arange(0, len(Yt), 1)
    x0 = [50,3,0.1]
    Z = kalman_filter_with_fitting(Yt, time, x0, sigma_o, start_index, t_b)

    profit = trading_algorithm(open_prices, close_prices, Z, start_index)
    print(sigma_o, t_b, profit[len(profit)-1])
    return profit[len(profit)-1]

sigma_o_values = np.logspace(-2, 2, 20)  
t_b_values = np.arange(20, 201, 30)  

profits = np.zeros((len(t_b_values), len(sigma_o_values)))

# Compute the profit for all combinations of sigma_o and t_b
for i, sigma_o in enumerate(sigma_o_values):
    for j, t_b in enumerate(t_b_values):
        profits[j, i] = profit_function([sigma_o, t_b], 
                                        open_prices[:len(open_prices)-252], 
                                        close_prices[:len(open_prices)-252], 
                                        start_index)

max_index = np.unravel_index(np.argmax(profits, axis=None), profits.shape)
max_profit_sigma_o = sigma_o_values[max_index[1]]
max_profit_t_b = t_b_values[max_index[0]]
max_profit_value = profits[max_index]

plt.figure(figsize=(10, 6), dpi=300)
plt.pcolormesh(sigma_o_values, t_b_values, profits, cmap='RdYlGn')
cbar = plt.colorbar(label='Profit [%]')  # Create colorbar
cbar.ax.yaxis.label.set_fontsize(20)
cbar.ax.yaxis.set_tick_params(labelsize=18) 
plt.xscale('log')
plt.xlabel(r'$\sigma_o$', size=20)
plt.ylabel(r'$t_b$', size=20)
plt.xticks(fontsize=18)
plt.yticks(fontsize=18)
plt.title('Algorithm profit over 3-year period', size=22)
plt.savefig("meshgrid-params.svg")
plt.show()
\end{python}

\subsection{Simulation of the Heston model}
\label{sec: Heston simulation}
\begin{python}
np.random.seed(seed=605)

#Model parameters
S = 100         # initial stock price
T = 1           # time to maturity
r = 0.05        # risk-free rate
kappa = 2.0     # speed of mean reversion
theta = 0.02    # long-term mean of the volatility
v_0 = 0.01      # initial volatility
rho = -0.7      # correlation between the Brownian motions
xi = 0.1        # volatility of the volatility
steps = 500     # number of steps
Npaths = 5      # number of paths

# Function to generate Heston model paths
def generate_heston_paths(S, T, r, kappa, theta, v0, rho, xi, steps, Npaths):
    dt = T / steps
    prices = np.zeros((Npaths, steps))
    volatilities = np.zeros((Npaths, steps))
    for i in range(Npaths):
        prices[i, 0] = S
        volatilities[i, 0] = v0
        for t in range(1, steps):
            dW1 = np.random.normal(0, np.sqrt(dt))
            dW2 = rho * dW1 + np.sqrt(1 - rho**2) * np.random.normal(0, np.sqrt(dt))
            prices[i, t] = prices[i, t-1] * np.exp((r - 0.5 * volatilities[i, t-1]) * dt + np.sqrt(volatilities[i, t-1]) * dW1)
            volatilities[i, t] = volatilities[i, t-1] + kappa * (theta - volatilities[i, t-1]) * dt + xi * np.sqrt(volatilities[i, t-1]) * dW2
    return prices

# Generate the price paths
prices = generate_heston_paths(S, T, r, kappa, theta, v_0, rho, xi, steps, Npaths)

# Plot the paths of the prices
plt.figure(figsize=(10, 5), dpi=300)
time_axis = np.linspace(0, T, steps)
for i in range(Npaths): 
    plt.plot(time_axis, prices[i], label=f'Stock Price Path {i+1}')
plt.title('Simulated Heston Model Price Paths',size = 20)
plt.xlabel('Time',size = 18)
plt.ylabel('Stock Price',size = 18)
plt.savefig("heston_simulation.svg")
plt.show()
\end{python}

\subsection{Finding the MOM estimators for stock open price}
\label{sec: code for MOM-OPEN}
\begin{python}
    price_ratios = open_prices / open_prices.shift(1)
    price_ratios = price_ratios.dropna()
    n = price_ratios.count()
    m1_open = price_ratios.sum()/ n
    mu_open = m1_open - 1
    print(mu_open)
    price_ratios_fourth = price_ratios ** 4
    m4_open = price_ratios_fourth.sum()/n
    price_ratios_fifth = price_ratios ** 5
    m5_open = price_ratios_fifth.sum()/n
    mu = mu_open
    theta = theta_open
    def equations(p):
    k, x = p
    eq1 = (1/(k*(k - 2))) * (k**2 * mu**4 + 4*k**2 * mu**3 - 2*k * mu**4 - 8*k * mu**3 + 6*k**2 * mu**2 * theta -
                             12*k * mu**2 * theta + 6*k**2 * mu**2 - 12*k * mu**2 + 12*k**2 * mu * theta -
                             24*k * mu * theta + 4*k**2 * mu - 8*k * mu + 3*k**2 * theta**2 - 6*k * theta**2 -
                             3 * x**2 * theta + k**2 - 12*k * theta - 2*k) - m4_open

    eq2 = (1/(k*(k - 2))) * (k**2 * mu**5 + 5*k**2 * mu**4 + 10*k**2 * mu**3 * theta - 2*k * mu**5 +
                             10*k**2 * mu**3 + 30*k**2 * mu**2 * theta + 15*k**2 * mu * theta**2 -
                             10*k * mu**4 - 20*k * mu**3 * theta + 10*k * mu**2 * theta**2 + 30*k * mu * theta +
                             15*k**2 * theta**2 - 20*k * mu**3 - 60*k * mu**2 * theta - 30*k * mu * theta**2 -
                             15 * x**2 * theta + 5*k**2 * mu + 10*k**2 * theta - 20*k * mu**2 - 60*k * mu * theta -
                             30*k * theta**2 - 15 * x**2 * theta + k**2 - 10*k * mu - 20*k * theta - 2*k) - m5_open

    return (eq1, eq2)
    v = (mu_open, theta_open)  
    kappa, xi = fsolve(equations, v)
    print(f"Solved kappa: {kappa}")
    print(f"Solved xi: {xi}")

\end{python}

\subsection{Finding the MOM estimators for stock close price}
\label{sec: code for MOM-CLOSE}
\begin{python}
    price_ratios = close_prices / close_prices.shift(1)
    price_ratios = price_ratios.dropna()
    n = price_ratios.count()
    m1_close = price_ratios.sum()/ n
    mu_close = m1_close - 1
    price_ratios_sqr = price_ratios ** 2
    m2_close = price_ratios_sqr.sum()/ n
    theta_close = m2_close - (1 + mu_close)**2
    price_ratios_fourth = price_ratios ** 4
    m4_close = price_ratios_fourth.sum()/n
    price_ratios_fifth = price_ratios ** 5
    m5_close = price_ratios_fifth.sum()/n
    mu = mu_close
    theta = theta_close
    def equations(p):
    k, x = p
    eq1 = (1/(k*(k - 2))) * (k**2 * mu**4 + 4*k**2 * mu**3 - 2*k * mu**4 - 8*k * mu**3 + 6*k**2 * mu**2 * theta -
                             12*k * mu**2 * theta + 6*k**2 * mu**2 - 12*k * mu**2 + 12*k**2 * mu * theta -
                             24*k * mu * theta + 4*k**2 * mu - 8*k * mu + 3*k**2 * theta**2 - 6*k * theta**2 -
                             3 * x**2 * theta + k**2 - 12*k * theta - 2*k) - m4_open

    eq2 = (1/(k*(k - 2))) * (k**2 * mu**5 + 5*k**2 * mu**4 + 10*k**2 * mu**3 * theta - 2*k * mu**5 +
                             10*k**2 * mu**3 + 30*k**2 * mu**2 * theta + 15*k**2 * mu * theta**2 -
                             10*k * mu**4 - 20*k * mu**3 * theta + 10*k * mu**2 * theta**2 + 30*k * mu * theta +
                             15*k**2 * theta**2 - 20*k * mu**3 - 60*k * mu**2 * theta - 30*k * mu * theta**2 -
                             15 * x**2 * theta + 5*k**2 * mu + 10*k**2 * theta - 20*k * mu**2 - 60*k * mu * theta -
                             30*k * theta**2 - 15 * x**2 * theta + k**2 - 10*k * mu - 20*k * theta - 2*k) - m5_open

    return (eq1, eq2)
    v = (mu_close, theta_close)  
    kappa, xi = fsolve(equations, v)
    print(f"Solved kappa: {kappa}")
    print(f"Solved xi: {xi}")
    
\end{python}

\subsection{Find the optimal $\rho$ and calculation of the sum of absolute error of estimation of the stock open price}
\label{sec: optimal rho and sum of ae-3yr}
\begin{python}
    np.random.seed(605)
    theta = theta_open  
    xi = xi
    k = kappa      
    n = n      
    V_0 = 0      
    V_t = np.zeros(n+1)
    V_t[0] = V_0
    Z_1 = np.random.normal(size=n)
    for t in range(1, n+1):
    V_t[t] = V_t[t-1] + k * (theta - V_t[t-1]) + xi * np.sqrt(np.abs(V_t[t-1])) * Z_1[t-1]
    V_t[:10]  
    n = len(open_prices)  
    Z = np.random.normal(size=2*n)
    Z_1 = Z[:n]
    Z_2 = Z[n:]
    S_t = np.zeros(n+1)
    S_t[0] = first_day_open_price
    mu_open = mu_open  
    V_t = V_t
    def calculate_S_t(rho):
    for t in range(1, n+1):
        S_t[t] = (1 + mu_open + np.sqrt(np.abs(V_t[t-1])) * (rho * Z_1[t-1] + np.sqrt(1 - rho**2) * Z_2[t-1])) * S_t[t-1]
    return S_t
    best_rho = None
    min_diff_sum = np.inf
    for rho in np.arange(-1, 1, 0.001):  
    S_t = calculate_S_t(rho)
    diff_vec = np.abs(open_prices - S_t[1:])  
    diff_sum = np.sum(diff_vec)
    if diff_sum < min_diff_sum:
        min_diff_sum = diff_sum
        best_rho = rho
    print(f"The best rho is {best_rho} with a minimum difference sum of {min_diff_sum}")

\end{python}

\subsection{Find the optimal $\rho$ and calculation of the sum of absolute error of estimation of the stock close price}
\label{sec: optimal rho and sum of ae-3yr-close}
\begin{python}
    np.random.seed(605)
    theta = theta_close  
    xi = xi
    k = kappa      
    n = n      
    V_0 = 0      
    V_t = np.zeros(n+1)
    V_t[0] = V_0
    Z = np.random.normal(size=2*n)
    Z_1 = Z[:n]
    for t in range(1, n+1):
    V_t[t] = V_t[t-1] + k * (theta - V_t[t-1]) + xi * np.sqrt(np.abs(V_t[t-1])) * Z_1[t-1]
    V_t[:10]  
    first_day_close_price = close_prices.iloc[0]
    first_day_close_price 
    n = len(open_prices)  
    Z = np.random.normal(size=2*n)
    Z_1 = Z[:n]
    Z_2 = Z[n:]
    S_t = np.zeros(n+1)
    S_t[0] = first_day_close_price
    mu_close = mu_close  
    V_t = V_t
    def calculate_S_t(rho):
    for t in range(1, n+1):
        S_t[t] = (1 + mu_close + np.sqrt(np.abs(V_t[t-1])) * (rho * Z_1[t-1] + np.sqrt(1 - rho**2) * Z_2[t-1])) * S_t[t-1]
    return S_t
    best_rho = None
    min_diff_sum = np.inf
    for rho in np.arange(-1, 1, 0.001):  
    S_t = calculate_S_t(rho)
    diff_vec = np.abs(close_prices - S_t[1:])  
    diff_sum = np.sum(diff_vec)
    if diff_sum < min_diff_sum:
        min_diff_sum = diff_sum
        best_rho = rho
        print(f"The best rho is {best_rho} with a minimum difference sum of {min_diff_sum}")
        
\end{python}



\end{document}